\documentclass{aa}         
\usepackage[utf8]{inputenc}
\usepackage{textcomp}
\usepackage{graphicx}      
\usepackage{hyperref}       
\usepackage{natbib}        
\bibpunct{(}{)}{;}{a}{}{,}  
\usepackage{pgfplotstable} 
\usepackage{booktabs}      
\usepackage{longtable} 
\usepackage{supertabular} 
\usepackage{amsmath}	
\usepackage{pifont} 
\usepackage{subcaption}
\usepackage{float} 
\usepackage{stfloats} 
\usepackage{placeins}
\usepackage{xcolor} 
\usepackage[normalem]{ulem}

\begin{document}

\title{First spectroscopic identification of the main sequence in Westerlund  1}
\titlerunning{IR main sequence of Westerlund 1}

\author{R.Castellanos \inst{1,}\inst{2}, 
F.Najarro\inst{1},
M.Garcia\inst{1},
I.Negueruela\inst{3},
L.R.Patrick\inst{1},
B.Ritchie\inst{4},
M.G.Guarcello\inst{5},
T.Shenar \inst{6},
C. Evans \inst{7},
R.Prinja \inst{8},
\and D. Fenech \inst{9}}
\authorrunning{Castellanos et al.}       

\institute{
    Departamento de Astrofísica, Centro de Astrobiología (CSIC-INTA), Ctra. Torrejón a Ajalvir km 4, E-28850 Torrejón de Ardoz, Spain 
    \and
    Departamento de Física Teórica, Universidad Autónoma de Madrid (UAM), Campus de Cantoblanco, E-28049 Madrid, Spain
    \and
    Departamento de Física Aplicada, Facultad de Ciencias, Universidad de Alicante
    \and
    School of Physical Sciences, The Open University, Walton Hall,Milton Keynes MK7 6AA, UK
    \and
    Istituto Nazionale di Astrofisica (INAF) – Osservatorio Astronomico di Palermo, Piazza del Parlamento 1, 90134 Palermo, Italy
    \and
    The School of Physics and Astronomy, Tel Aviv University, Tel Aviv 6997801, Israel
    \and 
    European Space Agency (ESA), ESA Office; Space Telescope Science Institute, 3700 San Martin Drive; Baltimore, MD 21218, USA
    \and
    Department of Physics and Astronomy, University College London, Gower Street, London WC1E 6BT, UK
    \and
    Square Kilometre Array Observatory (SKAO), Jodrell Bank, Lower Withington, Macclesfield SK11 9FT, UK
}

\date{Received 13 November 2025; accepted 24 February 2026}

\abstract
{Being the most massive known young stellar cluster in the Milky Way, Westerlund 1 (Wd1) constitutes an ideal benchmark for understanding the evolution of massive stars.
However, the cluster age remains highly controversial ($\sim$4-10 Myr), hindering the use of Wd1 as a reference for massive star evolution. One of the main issues is high foreground extinction, which has so far prevented the detection of the main sequence.
}
{Using infrared spectroscopy we seek to detect the cluster’s main sequence for the first time, to characterise the Hertzsprung-Russell diagram, and to use the cluster's turn-off to obtain a robust age estimate.}
{We obtained multi-epoch, near-infrared VLT/KMOS spectroscopic observations of Wd1 to map its population of massive stars. The spectra of $\sim$110 members were analysed with CMFGEN models to derive stellar parameters, populate the cluster Hertzsprung–Russell diagram, and compare it with isochrones from evolutionary models.}
{Our observations returned 47 new spectroscopically identified cluster members, with spectral types O9–B1 III–V. 
The cluster turn-off indicates an age of $5.5\pm1.0$ Myr at a distance of $4.23^{+0.23}_{-0.21}$\,kpc, displaying a moderate degree of coevality. 
We demonstrate that our estimate of the age of Wd1 is robust against reasonable changes in the distance and extinction law, and the adopted rotational velocity and metallicity of the stellar isochrones.
We further find that $\sim$65\% of the OB stars with multi-epoch coverage exhibit radial-velocity variability.}
{Infrared observations of the unevolved stellar population support a single episode of star formation
with an age of $\sim$5.5 Myr, reinforcing its potential as a benchmark for massive star evolution and providing a reference sample for future binary population studies.}

\keywords{Open clusters and associations: individual: Westerlund 1 - stars: evolution – stars: early-type}

\maketitle

\section{Introduction}\label{sec:intro}

Wd1 is renowned as one of the most massive young clusters in our Galaxy. 
Its total stellar mass is estimated on the order of $10^5\,M_\odot$ \citep{Clark2005}, and it hosts an extraordinary population of evolved high-mass stars. 
Notably, Wd1 contains about two dozen Wolf–Rayet (WR) stars, multiple red and yellow supergiants/hypergiants (RSG, YSG/YHG), at least one luminous blue variable (LBV), and an assortment of OB supergiants.
In addition, it harbours an X-ray pulsar that likely originated from a progenitor with a mass greater than $40M_\odot$ \citep{Muno2006}.
This rich stellar census makes Wd~1 a unique laboratory for studying the evolution of very massive stars and the formation of compact end-products in a dense cluster environment.

Despite its importance, Wd1 has been historically challenging to observe and interpret. 
Discovered by \citet{Westerlund1961}, Wd1 remained largely unstudied for decades due to foreground extinction ($A_V \approx 11\text{--}13\,$mag), which rendered it practically inaccessible to early optical surveys \citep{Negueruela2010, Damineli2016}. 
Reaching down to a magnitude of $V \approx 20$, \citet{Clark2005} still detected only evolved, luminous stars, with no trace of the cluster’s main sequence members.

Infrared observations have been essential to overcome these limitations.
By operating at wavelengths less affected by dust, near-infrared (NIR) surveys began to unveil Wd1’s hidden members that optical observations missed. 
NTT/SOFI JHK spectroscopy of the bright evolved WR population, allowed the classification of these objects \citep{Crowther2006}.
\citet{Brandner2008} detected the intermediate- and low-mass population of Wd1 using deep NIR photometry, while recent James Webb Space Telescope (JWST) observations have even begun to identify sub-solar mass members and protostellar disks by peering through Wd1’s dust \citep{Guarcello2024}. 
Complementary to these studies, the EWOCS (Extended Westerlund 1 and 2 Open Clusters Survey) collaboration is 
conducting a coordinated optical, infrared and X-ray campaign to extend the census from sub-stellar objects up to the most luminous WR stars \citep{Guarcello2024I,Anastasopoulou2024}.

This collection of increasingly comprehensive optical and IR spectroscopy has laid the groundwork for quantifying Wd\,1’s stellar content and fundamental parameters.  
An initial 4--5\,Myr age estimate was derived from the coexistence of $\sim$24 WR stars with four red supergiants and a substantial population of yellow/blue hypergiants \citep{Crowther2006, Negueruela2010}.
Optical spectroscopy confirmed $\sim$200 bright post main sequence cluster members \citep{Clark2005}, and subsequent surveys in the $I-$ and $R-$ bands have filled out the luminous census, revealing a smooth sequence from O9\,III giants to B\,I supergiants and finding no evidence for an older population \citep{Clark2020}.  
These data reinforced the interpretation of Wd1 as a roughly coeval ensemble about 5 Myr old, with binary interactions likely playing an important role in explaining the cluster’s large cohort of hypergiants \citep{Clark2011,Clark2020}.

However, recent studies have cast doubt on the $\sim$5 Myr age.
Applying a revised, steep optical--IR reddening curve specific to the cluster sightline \citep{Damineli2016}, \citet{Beasor2021} found that Wd\,1’s cool supergiants are $\sim$0.4\,dex too faint for a 5\,Myr isochrone, favouring a $\sim$10 Myr age.  
While age estimates based on the relative numbers of evolved stars were independent of distance and extinction, analyses relying on luminosities or HRD positions are highly sensitive to these parameters. 

Recent distance estimates in the literature range from $d\simeq2.8$ to $5$\,kpc.  
A Bayesian analysis of the Gaia EDR3 parallaxes yields $2.8^{+0.7}_{-0.6}$ kpc \citep{Aghakhanloo2021}.
The location of the pre-main sequence in the NIR colour-magnitude diagram yields $3.6$\,kpc \citep{Brandner2008} and $3.8$\,kpc \citep{Lim2013}.  
Independent constraints are derived from eclipsing binaries:
The modelling of W13 binary system yields $3.7\pm0.6$ kpc \citep{KoumpiaBonanos2012}, and a proper-motion–selected sample of OB binaries in \textit{Gaia} DR2 $3.87^{+0.95}_{-0.64}$ kpc \citep{DaviesBeasor2019}.  
Based on \textit{Gaia}~EDR3 proper motions and an improved parallax zero point, \citet{Negueruela2022} derived a distance of $d = 4.23^{+0.23}_{-0.21}$\,kpc, in better agreement with early spectro-photometric studies that favoured $\sim$5\,kpc \citep{Clark2005,Crowther2006} when systematic effects are considered.
The sightline extinction, meanwhile, spans $A_{V}\!\simeq\!10$--12\,mag with marked spatial variation \citep{Andersen2017}.  
Considering the outstanding discrepancies in distance and reddening, the age of Wd1 remains unsettled.

The motivation for our present study is to overcome the observational limitations that have so far restricted our understanding of Wd1. 
We seek to peer through the heavy dust extinction that hinders optical surveys and capture the NIR spectra of Wd1’s main sequence massive stars for the first time.
Our objectives are to assess the full cluster population, derive a refined age for Wd1, clarify its evolutionary state, thereby resolving the current age and distance discrepancies and cementing Wd1’s role as a cornerstone of massive-star evolution. 
In the following sections we describe the observations and data reduction (Sect.~\ref{sec:Observations}), the spectral analysis and stellar-parameter determination (Sect.~\ref{sec:Stellar_parameters}), the construction of the HR diagram and age estimate (Sect.~\ref{sec:HRD}), discuss the broader implications (Sect.~\ref{sec:discussion}), and summarise our main conclusions (Sect.~\ref{sec:conclusions}).

\section{Observations and data reduction}
\label{sec:Observations}

\subsection{VLT/KMOS observations}

Spectroscopic observations were obtained with the K-band Multi Object Spectrograph (KMOS; \citealt{Sharples2013}) mounted on ESO’s Very Large Telescope (VLT), through two separate observing programmes: a single-epoch campaign during Period~109 (P109, P.I.: R.~Castellanos), and a multi-epoch follow-up in Period~113 (P113, P.I.: R.~Castellanos). 
Both campaigns employed similar instrumental setups and observational strategies, allowing their combination into a uniform dataset. 

KMOS's 24 deployable Integral Field Units (IFUs) allowed simultaneous observation of multiple targets within the $\sim$4\arcmin\ field of view.
The IFUs were allocated to the program candidates (Sect. \ref{sec:Sample}) avoiding source crowding.

In P109, a total of $\sim$150 stars were observed across 10 individual KMOS configurations. 
Each configuration was observed once, with spectroscopy acquired in the $YJ$-, $H$-, and $K$-bands, with spectral resolving power of R($\lambda$/$\Delta$$\lambda$)~$\sim$~3400, 3900, and 4200, respectively.

In P113, a subsample of $\sim$90 stars was subject to further observation using five different KMOS configurations. 
Each configuration was repeated up to five times in order to enable the detection of radial velocity variability indicative of binarity.
To maximise the signal-to-noise ratio as NIR spectral lines are intrinsically weak in main sequence stars, this campaign was restricted to the $YJ-$ and $H-$ bands.

Observations in both programmes were conducted in service mode under excellent seeing conditions ($<$0.8\arcsec), using an ABA dither pattern with offset sky positions for optimal background subtraction. 
Exposure times were tailored to source brightness, with detector integration times (DITs) ranging from 6~s to 196~s per exposure, and 4--5 exposures per grating. 

A summary of the observational setup and parameters for both programmes is provided in Table~\ref{tab:obs_summary}.

\begin{table*}
\caption{Summary of the VLT/KMOS observing campaigns.}
\label{tab:obs_summary}
\centering
\small       % smaller font for large tables
\renewcommand{\arraystretch}{1.2}   % vertical padding
\begin{tabular}{lccccc}
\hline\hline
Programme & Period & Dates & Configurations & Epochs & Bands / DIT (s)\\
\hline
109.233D.001 & 109 & 2022 Apr--Jun & 10 & 1 & $YJ,H,K$\,/\,6–117 \\
113.26BB.001 & 113 & 2024 Apr--Jul & 5  & 5 & $YJ,H$\,/\,8–196 \\
\hline
\end{tabular}
\end{table*}

\subsection{Sample selection} \label{sec:Sample}
Targets were selected from the NIR photometric catalogue of \citet{Gennaro2011}, based on SOFI/NTT observations, which provides $JHK$ photometry suitable for identifying both bright evolved stars and fainter unevolved candidates.
From the $\sim$150 KMOS spectra obtained, we discarded stars for whose signal-to-noise ratio (S/N) was not sufficient (typically below 100) to reliably measure the weak diagnostic lines available in the observed bands. 
As we were mainly interested in the less evolved population of the cluster in this work, we excluded most of the known WR stars as well as double-lined spectroscopic binaries.
This resulted in a sample of 109 stars to be analysed in this work (see Table~\ref{tab:Stellar_parameters} and Fig.~\ref{fig:HRD}).

The sample includes two main groups. 
Approximately 60 stars correspond to previously known cluster members that have been spectroscopically classified at optical wavelengths, primarily from the VLT/FLAMES survey by \citet{Clark2020}.
These objects constitute a reference population of evolved massive stars.

The remaining $\sim$50 stars had been selected as candidate main sequence members using photometric criteria designed to isolate early-type OB stars behind the strong extinction affecting Wd1.
Specifically, we applied the following selection criteria, $J > 12$~mag and $1.2 < J-K < 1.5$, guided by the observed magnitudes of previously classified supergiants and giants in the cluster.
They were further chosen to minimise contamination from foreground sources and evolved stars.

\subsection{Data reduction and telluric correction}

All spectroscopic data were reduced using the standard \texttt{esoreflex} KMOS pipeline (SPARK, \citealt{Davies2013}) in the ESO Reflex automated data reduction environment \citep{Freudling2013}.
This workflow includes flat-field correction, wavelength calibration, bad pixel masking, cube reconstruction, and extraction of one-dimensional spectra from the IFUs. 
The same reduction strategy was applied to both the single-epoch (P109) and multi-epoch (P113) observations to ensure consistency across the dataset.
A critical step in the reduction process was the correction of telluric absorption features, particularly severe in the near-infrared due to strong water vapour bands. 
Two methods were tested for telluric correction: the classical standard-star approach and the Molecfit software.

In the first approach, early-type standard stars (preferentially A0V) were observed at airmasses similar to those of the science targets, either immediately before or after the science observations. 
These standards, with nearly featureless continuum in the NIR, were used to derive telluric transmission spectra after modelling and removing their intrinsic stellar lines.
These spectra were then used to remove the telluric absorption from the science spectra. 

The second approach makes use of Molecfit \citep{Smette2015}, which fits synthetic atmospheric transmission models directly to the observed spectra. 
This method has proven highly effective for VLT/KMOS data, especially in cases where sky conditions were variable or the quality of standard-star observations was limited. 

\begin{figure}[!t]
    \centering
    \includegraphics[width=\columnwidth]{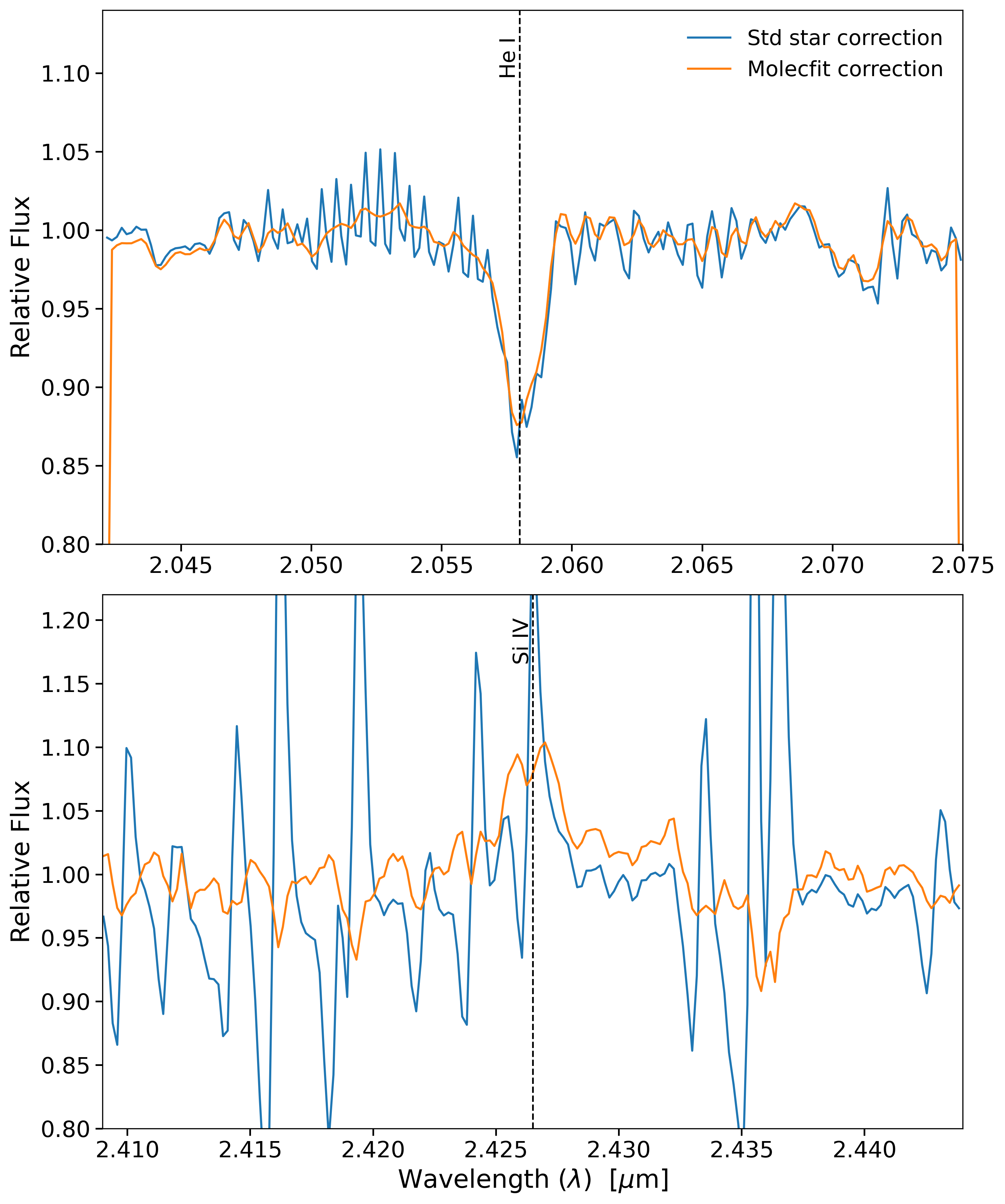}
    \caption{Comparison of the final, telluric-corrected science spectra obtained with the two considered methods for telluric correction, one based on a standard star (blue) and the other using Molecfit (orange). 
    Two wavelength intervals strongly affected by telluric absorption are shown: top, 2.04–2.075\,$\mu$m around the \ion{He}{i} line at 2.058\,$\mu$m; bottom, 2.41–2.45\,$\mu$m around \ion{Si}{iv} at 2.4265\,$\mu$m.}
    \label{fig:telluric_comp}
\end{figure}

In all our tests, Molecfit consistently yielded lower residuals around key diagnostic lines and was therefore adopted for the final analysis. 
Fig.~\ref{fig:telluric_comp} compares the resulting corrected spectra using the two methodologies described.

\section{Spectroscopic analysis}
\label{sec:Stellar_parameters}

\subsection{Near\,$\mathit{IR}$ spectral classification}\label{sec:classif}

Spectral types were assigned through visual comparison against reference standards ranging from O9 to B2, encompassing all luminosity classes. 
The primary reference grid consists of standard stars established by \citet{Negueruela2024}.  
Archival near-IR VLT/X-SHOOTER spectra were retrieved, continuum-normalised, and convolved to KMOS resolution, enabling a direct comparison of their line profiles against our targets.
We also considered the medium-resolution atlas of OB stars by \citet{Hanson2005}, whose spectra cover the $H$ and $K$ bands.
Finally, the grid was complemented with additional VLT/X-SHOOTER spectra of O9–B2 stars classified in the optical range. 

\begin{figure*}[t]               
  \centering
  \includegraphics[width=\textwidth]{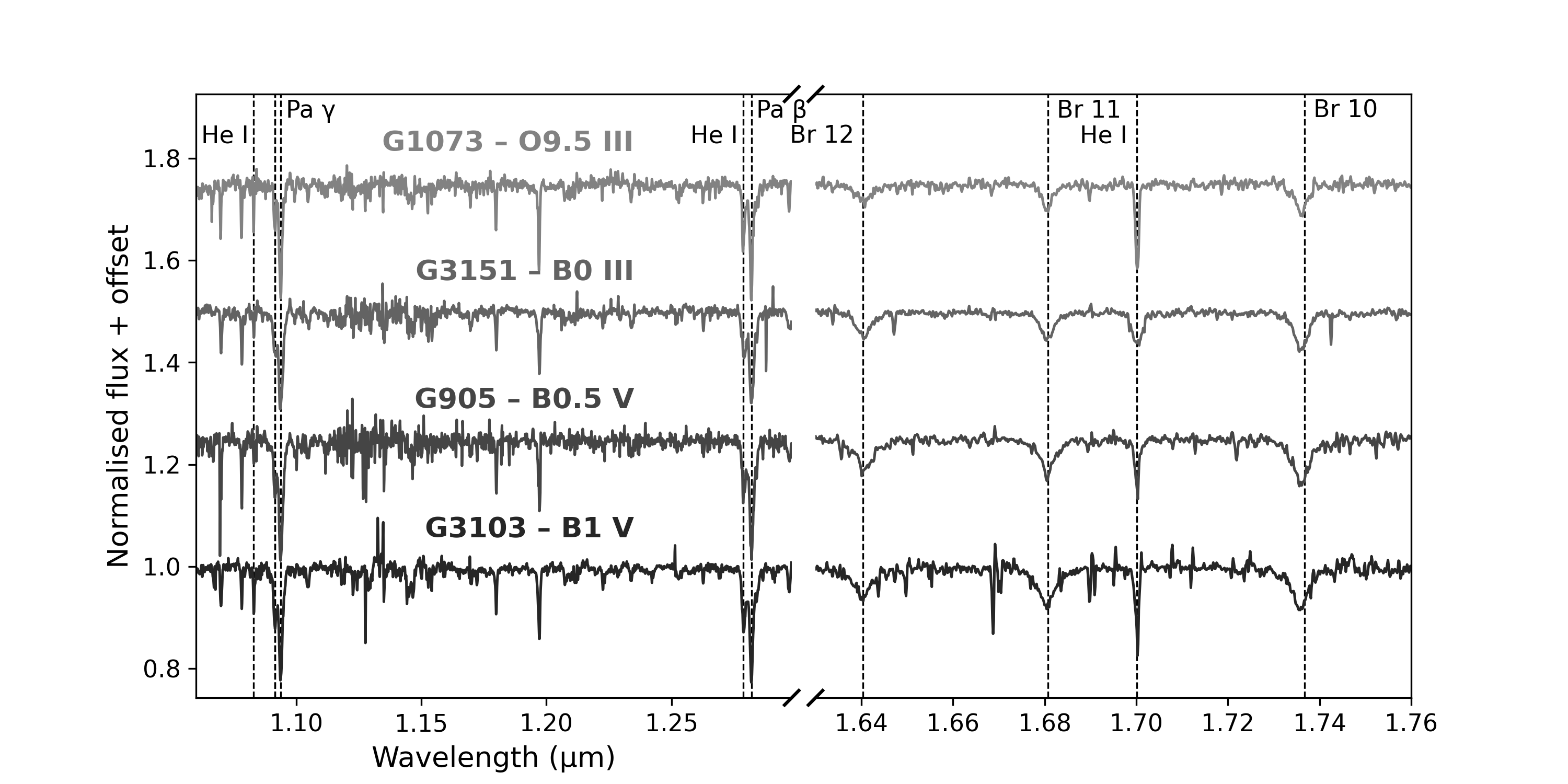}
  \caption{Normalised $J$- and $H$-band spectra of four newly classified OB members of Westerlund\,1, offset vertically for clarity. 
  Object IDs and adopted spectral types are indicated on the left.  
  Principal diagnostic lines used for the spectral classification include Pa\,$\delta$, 
  Pa\,$\gamma$,   
  He\,\textsc{i}\,1.083\,$\mu$m, He\,\textsc{i}\,1.092\,$\mu$m, He\,\textsc{i}\,1.278\,$\mu$m,
  Br\,10–12, 
  He\,\textsc{ii}\, 1.692\,$\mu$m, 
  and He\,\textsc{i}\,1.700\,$\mu$m.}  
  \label{fig:JH_examples}
\end{figure*}

The classification was performed mainly based on the observation of hydrogen and helium lines that exhibit variations with spectral subtypes in the infrared spectrum.
In the $J$ band we inspected the Paschen series (Pa\,$\delta$, Pa\,$\gamma$) together with the He\,\textsc{i}\,1.083\,$\mu$m, He\,\textsc{i}\,1.092\,$\mu$m and He\,\textsc{i}\,1.278\,$\mu$m lines.
Due to the variable spectral coverage across the KMOS IFUs, the He\,\textsc{ii}\,1.012\,$\mu$m line lies at the edge of the $J$-band range and is only detected in a few spectra, and thus was not used for classification purposes.
In the H band, classification relied on Br\,10–12, He\,\textsc{i}\,1.70\,$\mu$m and He\,\textsc{ii}\,1.69\,$\mu$m.
In the $K$ band we considered Br\,$\gamma$ alongside He\,\textsc{i}\,2.161\,$\mu$m, He\,\textsc{i}\,2.058\,$\mu$m, HeI 2.112/3\,$\mu$m doublet which overlaps with the C\,\textsc{iii}\,/N\,\textsc{iii}\,/O\,\textsc{iii}\, 2.115\,$\mu$m complex and Si\,\textsc{iv}\,2.426\,$\mu$m lines. 
Examples of the $J$- and $H$-band spectra of our sample are shown in Fig.~\ref{fig:JH_examples}.
Dwarf-type new members exhibit the deep and broad Brackett and Paschen profiles characteristic of this luminosity class.
Our new classification revisits the optical types previously assigned by \citet{Clark2020} for $\sim$60 stars, confirming the overall reliability of their scheme while refining several cases.
In most instances, the NIR classifications derived here agree within half a subtype with the optical types proposed by \citet{Clark2020}.

The $\sim$50 newly classified objects have been appended to the known cluster census (Table~\ref{tab:wd1_demo}) and, together with the OB members described in Sect.~\ref{sec:Sample}, form the final sample for which fundamental stellar parameters are derived in Sect.~\ref{sec:CMFGEN_grid}.
The assigned spectral types are listed in Table \ref{tab:Stellar_parameters}.

\begin{table}
  \caption{Stellar demographics of Wd1 updated from \citep{Clark2020}.}
  \label{tab:wd1_demo}
  \centering
  \begin{tabular}{lcc}
    \hline\hline
    Spectral   & Total population & New to \\
    type & (Included in this work) & this work \\
    \hline
    O9--9.5\,Iab,\,Ib            & 18 (6) & 0 \\
    O9--9.5\,II,\,II--III        & 57 (47) & 12 \\
    O9--9.5\,IV - V              & 9 (9) & 8 \\
    B0--0.5\,Ia,\,Iab,\,Ib       & 20 (8)  & 1  \\
    B0--0.5\,II,\,II--III       & 19 (19) & 9 \\
    B0--0.5\,IV - V              & 14 (14) & 14 \\
    B1--1.5\,Ia,\,Iab            & 9 (2) & 0  \\
    B1--1.5\,IV - V              & 3 (3) & 3 \\
    B2--4\,Ia                    & 7 (1)  & 0  \\
    \hline
    O4--8\,Ia$^{\ast}$           & 2 (0)   & 0  \\
    B0--2\,Ia$^{\ast}$/WNVL      & 4 (0)  & 0  \\
    B5--9\,Ia$^{\ast}$           & 4 (0)  & 0  \\
    \hline
    LBV                          & 1 (0)  & 0  \\
    YHG\,+\,RSG                  & 10 (0) & 0  \\
    \hline
    sgB[e]                       & 1 (0)  & 0  \\
    OB\,SB2                      & 12 (0) & 0 \\
    OeBe                         & 1 (0)  & 0  \\
    \hline
    WN5--8                       & 14 (0) & 0  \\
    WC8--9                       & 8 (0)  & 0  \\
    \hline
    \textbf{Total}               & \textbf{213 (109)} & \textbf{47} \\
    \hline
  \end{tabular}
\end{table}

\subsection{Radial velocity analysis}\label{sec:rv_analysis}

\begin{figure*}
  \centering
  \includegraphics[width=\linewidth]{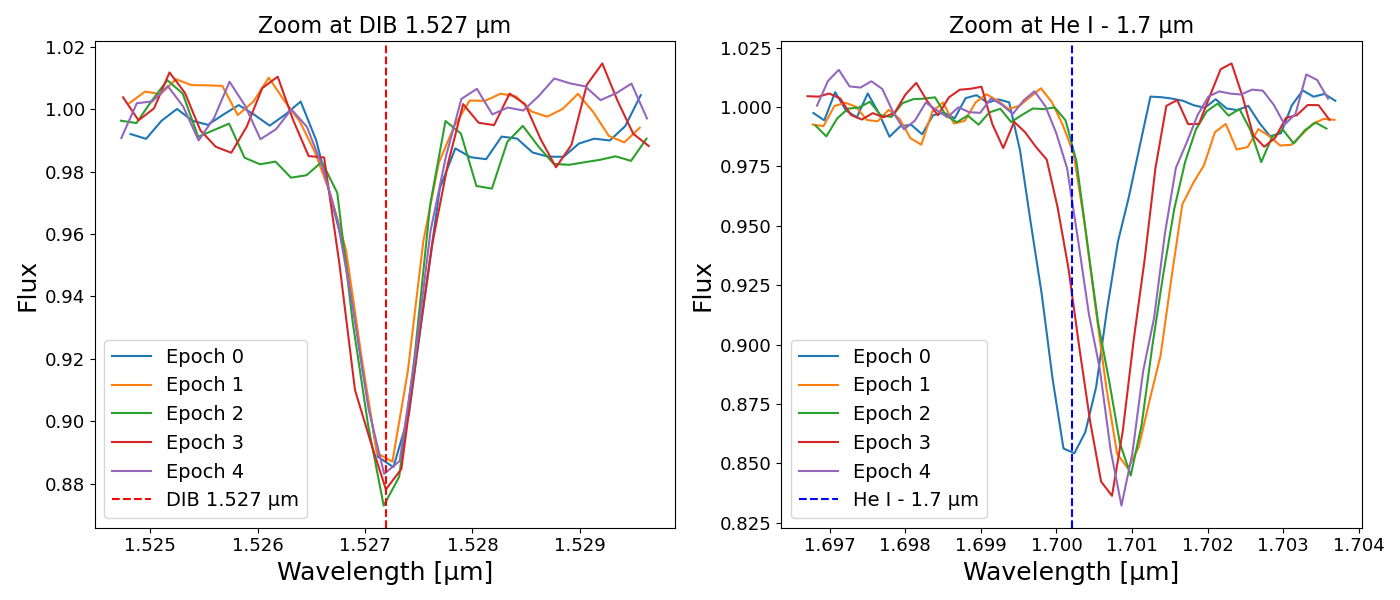}
  \caption{Zoom over the spectra used for the radial‐velocity measurement of Wd1-1048.  
  Coloured lines correspond to epochs as indicated in the legend.
  Left: DIB\,1.527 µm absorption in the five epochs, adopted as wavelength reference (vertical red dashed line marks the rest frame wavelength). 
  Right: He\,\textsc{i}\,1.70 µm line in the five epochs after realigning the spectra.
  Wd1-1048 is a known RV variable and possible eclipsing binary \citep{Ritchie2022}.}
  \label{fig:zoom_DIB_HeI}
\end{figure*}

Our multi-epoch KMOS observations enable a first-order search for spectroscopic binaries by monitoring radial-velocity (RV) changes between different observing epochs.
J- and H-band spectra were aligned in wavelength using NIR diffuse interstellar bands (DIBs) available in both regions. 
In particular, the DIB at 1.317 µm in J-band and the one at 1.527 µm in H-band were fitted with a Gaussian profile in each epoch. 
Individual spectra were then shifted to place the DIB centroid at a common rest wavelength as shown in Figure~\ref{fig:zoom_DIB_HeI}. 

Once aligned, relative radial velocity variations were measured from the Doppler shift of the strong stellar absorption lines He I 1.700 µm line in H-band.
The He I diagnostic lines in the J-band were discarded because they experience higher noise, and are contaminated by DIBs in some cases. 
Gaussian profiles were fitted to the line cores in each epoch, deriving the central wavelengths.

We flag candidate spectroscopic binaries in Table~\ref{tab:Stellar_parameters} if the following criteria are fulfilled:
(i) the spectra display a peak-to-peak variation exceeding $\Delta{\rm RV}>20$~km\,s$^{-1}$ following previous work on OB star surveys (e.g., \citet{Sana2013}) ;  
(ii) the variation reaches a significance level of $\Delta{\rm RV}\geq 4\sigma$ \citep{Bodensteiner2021}.  
We find that $63.1\%$ of the OB stars with multi-epoch coverage are flagged as RV variable.
This fraction exceeds that reported from optical FLAMES data \citep{Ritchie2022}
\footnote{Adopting a more conservative threshold of $\Delta{\rm RV}>25$~km\,s$^{-1}$, as in \citet{Ritchie2022}, would yield a detection fraction of $60.6\%$.}
which could be explained as follows: on the one hand, the optical sample is dominated by evolved supergiants, whose larger radii imply wider orbits and smaller RV amplitudes.
On the other hand, the optical sample could also include post-interaction systems.
Table~\ref{tab:Stellar_parameters} also includes the additional confirmed or candidate binaries compiled by \citet{Ritchie2022}.

A deeper analysis of binary aspects aiming at deriving the intrinsic binary fraction and placing constraints on the underlying period and mass-ratio distributions, will be presented in a forthcoming study.

\subsection{CMFGEN model fit}
\label{sec:CMFGEN_grid}

In order to derive the fundamental stellar parameters—effective temperature ($T_{\rm eff}$) and bolometric luminosity ($L_\star$)—for the observed sample, we fitted each observed spectrum against a precomputed grid of $\sim$5000 non-LTE, line-blanketed model atmospheres generated with the CMFGEN code \citep{HillierMiller1998}.
For stars with multi-epoch observations, the individual spectra were shifted to the same reference frame and co-added in order to increase the S/N before performing the CMFGEN fitting.

The CMFGEN grid utilised in this study samples the parameter space relevant for late O- and early B- type stars in Wd1. 
In practice, the grid encompasses effective temperatures ranging from 15 000 to 40 000 K with a nominal step of roughly 1 000 K.
Additionally, surface gravities are covered from $\log g=2.5$ to 4.0 (cgs) with a typical increment of approximately 0.15 dex, and including models computed for both solar and slightly metal-rich compositions.
Each model spectral energy distribution (SED) is computed from 0.3 to 5 µm, covering the wavelength range of the KMOS observations.

Each observed KMOS spectrum was first continuum normalised by fitting a low–order spline to carefully selected line-free regions across the $J-$, $H-$, and $K-$ bands.  
Using the He\,\textsc{i} $ 1.700\,\mu$m line in the $H$ band as reference, the projected rotational velocity ($v_{\sin i}$) 
and the radial–tangential macroturbulent velocity ($v_{\rm mac}$) 
were empirically constrained.
These parameters were varied in discrete steps and the combination providing the best agreement between a reference model and observations was adopted. 
These resulting values, together with the instrumental resolution, were then used to convolve all CMFGEN models.

We then compared each observed, normalised spectrum against the CMFGEN grid, by using the reduced chi squared metric:
\[
  \chi^2_\nu=\frac{1}{N_{\mathrm{pix}}-p}
             \sum_{i=1}^{N_{\mathrm{pix}}}
             \left[\frac{F_{\mathrm{obs}}(\lambda_i)-F_{\mathrm{mod}}(\lambda_i)}
                        {\sigma_i}\right]^2 ,
\]
where $F_{\mathrm{obs}}$ and $F_{\mathrm{mod}}$ are the observed and model fluxes, $\sigma_i$ is the per-pixel uncertainty, $N_{\mathrm{pix}}$ the number of wavelength points considered, and $p$ the number of free parameters.
The fit was restricted to windows that contain the main diagnostic features such as Pa\,$\gamma$ and Pa\,$\beta$ in the $J$ band; Brackett lines and He\,\textsc{i}\,$1.70\,\mu$m in the $H$ band; and Br\,$\gamma$ together with He\,\textsc{ii}\,$2.188\,\mu$m in the $K$ band, while regions contaminated by strong telluric residuals or poor S/N were masked out to avoid biases.

The model yielding the minimum $\chi^2_\nu$ was adopted as the best fit.  
This choice was subsequently visually validated by superposing the full $J\!+\!H\!+\!K$ synthetic spectrum on the data. 
The visual inspection confirmed the quality of the fit across all key lines and flagged composite or otherwise peculiar spectra (e.g.\ asymmetric profiles or excessive broadening) that might indicate unresolved binaries.
In order to estimate the statistical uncertainties we examined the $\chi^2$ distribution across the model grid. 
For each star, we computed the associated p-values and defined the 1$\sigma$ confidence range in $T_{\rm eff}$ and $\log g$.
This typically yielded uncertainties of roughly $\pm1000$\,K and  $\pm0.15$\, dex in $\log g$, depending on the spectral type and S/N ratio.

Representative examples of this procedure are provided in Appendix ~\ref{sec:Appendix_modelfit}, where we compare the observed key diagnostic lines of our targets with the best-fit CMFGEN model.

\subsection{Extinction correction and stellar luminosity}
\label{sec:ext-radius}

Stellar luminosities were calculated by rescaling the CMFGEN model SED until it reproduced the dereddened SOFI $JHK_s$ photometry. 
For this calculation, we adopted a distance of $4.23^{+0.23}_{-0.21}$\,kpc from \citet{Negueruela2022}, derived from Gaia EDR3 parallaxes of kinematically selected Wd1 members and carefully accounting for the foreground population along this sightline.
The uncertainty in distance propagates into a systematic uncertainty of $\simeq\pm 0.04$~dex in $\log L_\star$.
For every star, the best-fitting CMFGEN spectral energy distribution was convolved with the SOFI $JHK_s$ response curves to obtain intrinsic model magnitudes ($m^{\mathrm{syn}}_{J,H,K}$). 
The colour excess $E(J-K)$ was computed from the difference between the observed and intrinsic model colours. 
Using the extinction law of \citet{Hosek2018}, we then derived $A_K$ and obtained the corresponding dereddened magnitudes, finding a mean extinction of $A_K = 0.69 \pm 0.07$ mag, $A_H = 1.37 \pm 0.14$ mag and $A_J = 2.47 \pm 0.25$ mag.
The \citet{Damineli2016} extinction law would instead yield $A_K = 0.8\pm 0.08$, $A_H = 1.44\pm 0.15$, and $A_J = 2.62\pm 0.28$ mag.
The large standard deviation of our results reflects that internal reddening is not negligible in the cluster.
We also note the larger average extinction in the J-band compared to the \citet{Damineli2016} average $A_J = 2.34$ value.

After correcting the observed photometry for extinction, the stellar radius ($R_\ast$, defined at optical depth $\tau = 2/3$) was determined by scaling the model flux to match the dereddened $K$-band magnitude at the adopted distance. 
The corresponding bolometric luminosity $L_\ast$ was then obtained from the Stefan–Boltzmann law using the spectroscopic $T_{\mathrm{eff}}$. 
The resulting $T_{\mathrm{eff}}$ and $L_\ast$ values, listed in Table~\ref{tab:Stellar_parameters}, match well the expected parameters for the spectral types of the analysed stars.

Uncertainties in luminosity were derived by propagating the errors in $T_{\rm eff}$ from our analysis and accounting for those in the distance from \citet{Negueruela2022}.
The resulting uncertainties in $\log L_\star$ are typically $\sim$0.1 dex across the sample.
A detailed discussion on the impact of the adopted distance in the interpretation of the HRD is presented in Appendix~\ref{sec:Appendix_parameters}.
Further uncertainties may also arise from unresolved companions, which we also discuss in following sections.

\section{HR diagram}\label{sec:HRD}

\begin{figure*}  
  \centering
  \includegraphics[width=\linewidth]{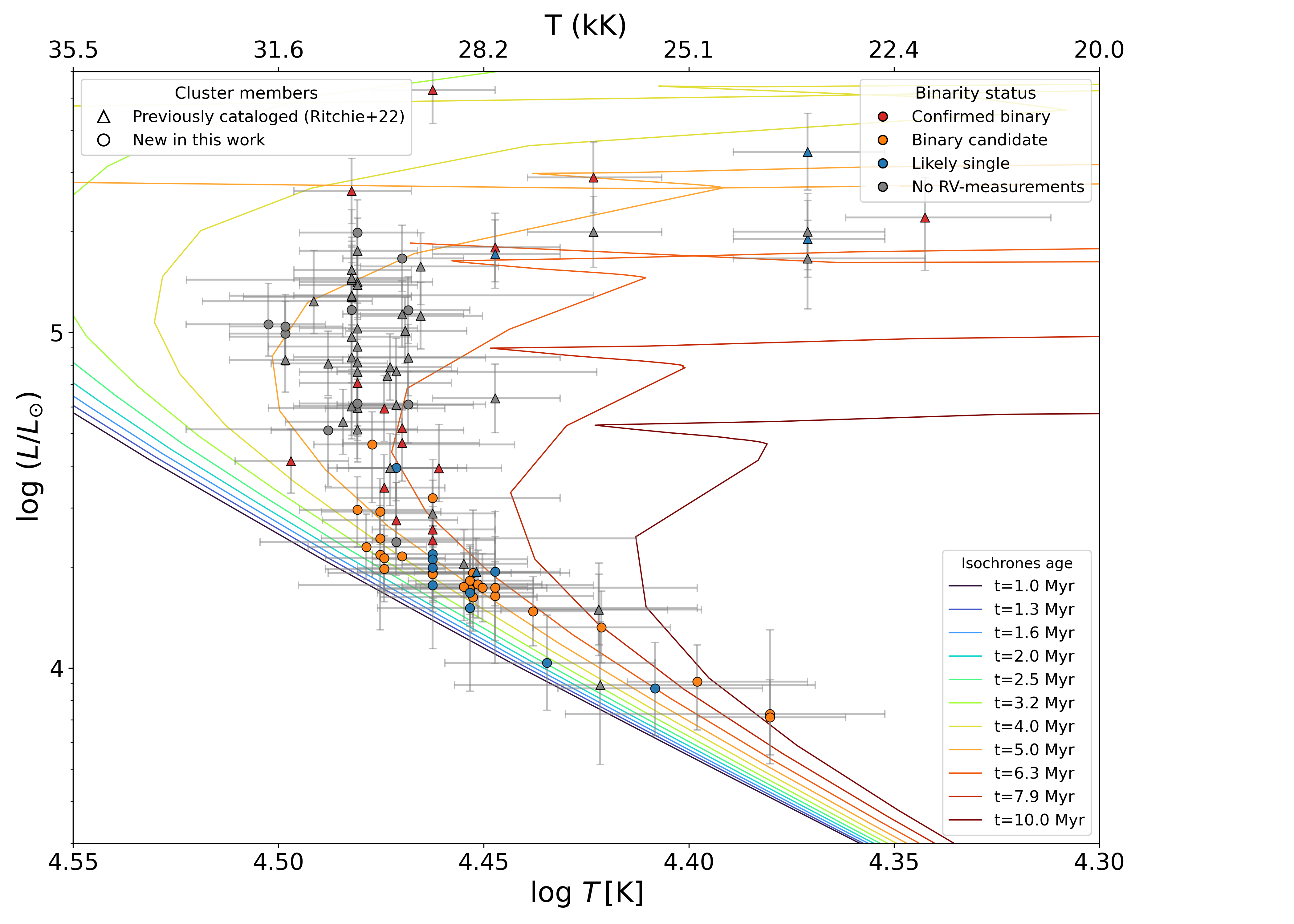}
  \caption{Hertzsprung–Russell diagram for the observed sample in Wd1.
  The colour of the symbols encodes the binary status based on RV measurements: 
  confirmed binaries from \citet{Ritchie2022} (red), 
  candidate binaries identified in this work (orange), 
  likely single stars (blue), 
  and stars without RV measurements (grey). 
  The shape of the symbols indicates whether the star had been analysed previously by \citet{Ritchie2022} (triangles),
  while circles mark stars analysed for the first time in this work. 
  $Z=0.02$ isochrones \citep{Yusof2022} for ages  $\log t\,[{\rm yr}] = 6.0$–7.0 are overplotted (coloured curves).}
  %\textcolor{red}{The luminosity error bars include the contribution from the uncertainty in the adopted cluster distance.}}
  \label{fig:HRD}
\end{figure*}

Using the $4.23^{+0.23}_{-0.21}$\,kpc distance value proposed by \citet{Negueruela2022} and the NIR extinction law of \citet{Hosek2018}, we construct the HR diagram for Wd1 with the stellar parameters compiled in Table \ref{tab:Stellar_parameters}. 
Figure~\ref{fig:HRD} shows the resulting HRD, where we overlay Geneva evolutionary isochrones for ages ranging from 1 to 10 Myr in 0.1-dex steps.  
An enriched metallicity of \(Z = 0.020\) was adopted, motivated by the elemental abundances derived for the cluster’s LBV Wd1-243 by \citet{Ritchie2009}.
The uncertainties on $T_{\rm eff}$ and $L_\star$, shown as error bars in Fig.~\ref{fig:HRD}, remain smaller than the spacing between neighbouring isochrones near the turn-off, ensuring a robust age determination within 0.1 dex.

The distribution of the cluster members reveals a well-defined main sequence, accompanied by the expected sequence of evolved giants and supergiants. 
The total number of considered stars is 109, of which 47 are newly identified O9-B0 III-V down to B1 V stars.
The remaining objects of the sample are previously known evolved giant and supergiant stars. 
The unevolved stars form a diagonal band in the HRD, spanning $T_{\rm eff}\sim25,000$–30,000~K and $\log(L_{\ast}/L_\odot)\sim~4.3$–4.8. 
The stars at the turn-off are spectroscopically classified as late-O to early-B giants, matching the spectral type for the predicted turn-off of an 5.5~Myr population: $T_{\rm eff} \approx 31,000$~K and $\log(L_\ast/L_\odot)~\approx~4.9$. 
Most evolved stars occupy slightly cooler temperatures ($\sim$23,000–27,000~K) and higher luminosities, consistent with the post-main-sequence tracks of $\sim$30–40~$M_\odot$~stars.

The cluster’s stars mostly gather between the 5–6.3~Myr isochrones.
In particular, the hottest giant stars are found near the 5 Myr track, and the luminous cool supergiants also coincide with the same age, strengthening a self-consistent single-population solution. 
The 5–6.3 Myr turn-off is a more suitable match than any older age.
For example, a 10 Myr isochrone would run significantly cooler and fainter than the observed main-sequence turn-off (by $\Delta T_{\rm eff}\sim$6000~K and $\Delta \log L_{\ast}\sim0.5$~dex).
Hence, we estimate a cluster age of $5.5\pm1.0$~Myr.
We do not find inconsistencies or signatures in the HRD that would suggest the presence of an additional, non-coeval population.
This point is further explored in Section~\ref{sec:discussion}.
We emphasize that only with the adopted distance of $4.23^{+0.23}_{-0.21}$~kpc do we achieve such consistency, as a shorter distance fails to reproduce the luminosities of both the unevolved and evolved members simultaneously.
We address this in detail in Appendix~\ref{sec:Appendix_parameters}.

We have also looked into possible biases in the position of stars due to multiplicity.
The RV variability of our targets (see Sect.~\ref{sec:rv_analysis}) is indicated in Fig.~\ref{fig:HRD}.
Stars lacking RV information are denoted by black triangles, while red symbols indicate previously identified variables from \citet{Ritchie2022}. 
Few of the RV variable stars in the main sequence lie slightly above the 6.3~Myr single-star isochrone on the HRD, consistent with unresolved binaries whose combined flux results in increased luminosity. 
If fluxes were de-blended, the corresponding points would shift to slightly lower luminosity.

To verify the 5.5$\pm$1.0~Myr age, we also consider the most evolved cluster members.
The presence of WRs in the cluster has been firmly established by previous studies (e.g., \citet{Clark2005}; \citet{Crowther2006}; \citet{Beasor2021} or \citet{Anastasopoulou2024}). 
Single stellar evolution theory predicts that, at $\sim$5.5~Myr, stars with initial masses above $\sim$40–50~$M_\odot$ will have already evolved into WRs or ended their lives \citep{Ekstrom2012, Yusof2022},consistent with the presence of a magnetar in the cluster reported by \citet{Muno2006}, while stars with slightly lower masses ($\sim$20–25 $M_\odot$) are expected to be entering the red supergiant phase. 
Our age estimate of $\sim$5.5$\pm$1.0~Myr at a distance of $4.23^{+0.23}_{-0.21}$~kpc remains compatible with the broader stellar population of the cluster, underpinning the age and fundamental parameters of Wd1. 

The impact of adopting different values for distance,extinction, metallicity and rotational velocity is explored in detail in Appendix~\ref{sec:Appendix_parameters}. 
We find that even when the largest plausible variations of these parameters are applied, the resulting shift is limited to a few tenths of a dex in $\log L_{\ast}$ and to less than 1 Myr in age, well within the adopted systematic uncertainties.

\section{Discussion}\label{sec:discussion}

Westerlund 1’s age ranges between $\sim$4~Myr and > 10~Myr in the literature.
Early analyses that relied exclusively on evolved stars consistently portrayed Wd1 as a very young cluster. 
\citet{Clark2005} noted the absence of O-type dwarfs in optical surveys and inferred an age of 4–5~Myr. 
The ratio of WR versus cool supergiants led \citet{Crowther2006} to a formal age of ${\sim}4.5$–5~Myr, and spectroscopy of $\sim$50 OB supergiants by \citet{Negueruela2010} likewise favoured a single, coeval population of at least 5 Myr.  
These studies portrayed Wd1 as a young, massive analogue to the Quintuplet cluster, with its abundant WR stars and first red supergiants fully compatible with a single episode of star formation, occurring ${\sim}5$ Myr ago, a picture that our NIR study now confirms.

More recent work challenged this view by focusing on the cool supergiant population.
\citet{Beasor2021} re-analysed the four red supergiants in Wd1 by integrating their observed spectral energy distributions, combining new SOFIA mid-infrared photometry with archival optical and NIR data. 
Their derived luminosities were too low to be reproduced by 5 Myr isochrones.
Instead, they argued that the RSGs are better matched by an older age, with a best-fit value of $10.4^{+1.3}_{-1.2}$~Myr, suggesting that Wd1 might be significantly older than previously thought, or alternatively, that it may host multiple stellar generations.
Under this interpretation, the progenitors required to form the $\sim$20 luminous WR stars observed in Wd1 should already have exploded, thus requiring a second, younger population to explain their presence in the cluster. 
Still, until now it was difficult to refute the older-age hypothesis because the main sequence had not been directly detected.

\begin{figure}[t]                      
  \centering
  \includegraphics[width=\columnwidth]{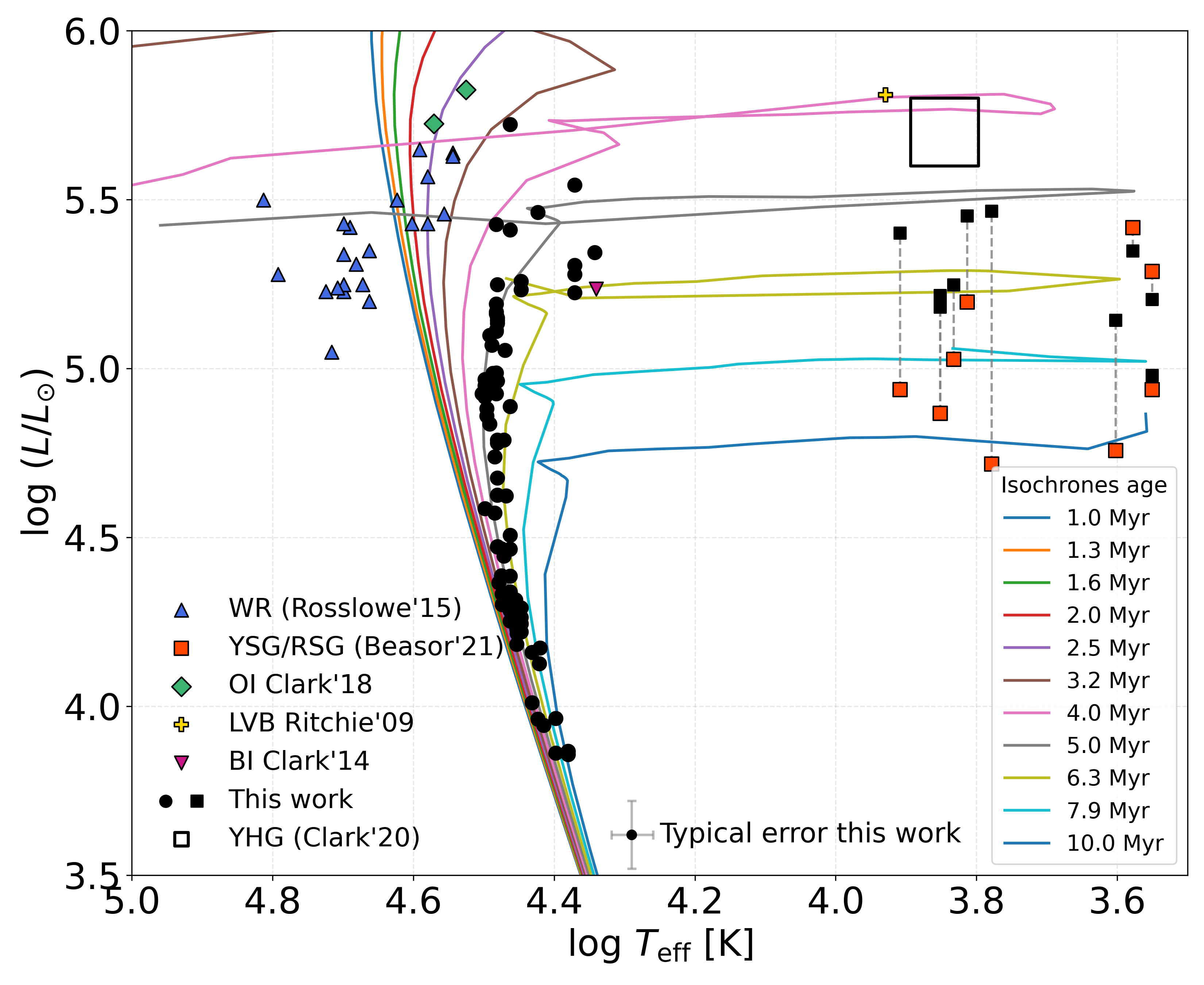}
  \caption{Hertzsprung–Russell diagram of Westerlund\,1 that combines the $\sim$110 OB stars analysed in this work (black circles) with selected evolved members from the literature.
  Representative error bars for the blue stars analysed in this work are provided at the bottom.
  Blue triangles denote WR stars \citep{Rosslowe2015}
  \protect\footnotemark\
  orange squares mark yellow and red supergiants  \citep{Beasor2021}; 
  black squares show the YSG and RSG luminosities recalculated in this work (Sect.~\ref{sec:YSGs}); 
  grey dashed lines connect each star to its value given in \citet{Beasor2021}, illustrating the offset between both approaches;
  green diamonds correspond to the blue-straggler hypergiants Wd1-27 and Wd1-30a \citep{Clark2019};
  the luminous-blue-variable Wd1-243 is shown as a yellow cross \citep{Ritchie2009};
  finally the supergiant Wd1-5 appears as a magenta inverted triangle \citep{Clark2014}.
  The region where YHGs would lie, following the luminosities proposed by \citet{Clark2020}, is indicated by an empty black square.
  All literature luminosities have been rescaled to the distance adopted here of $4.23^{+0.23}_{-0.21}$~kpc.
  Coloured curves are $Z=0.02$ isochrones \citep{Yusof2022} for ages between 1 and 10\,Myr, as indicated in the legend.}
  \label{fig:HR_combined}
\end{figure}

\footnotetext{We note that \citet{Rosslowe2015} used a different extinction law (based on \citet{Howarth1983} and \citet{Cardelli1989}) which may yield higher luminosities for the WR sample.}

The HRD in Fig.~\ref{fig:HR_combined} brings our KMOS OB sample together with all the evolved stars that have shaped the age debate, and for which stellar parameters have been published:
The full WR population \citep{Crowther2006,Rosslowe2015}, the RSGs and YSGs analysed by \citet{Beasor2021}, the blue-straggler hypergiants Wd1-27 and Wd1-30a \citep{Clark2019}, the extremely luminous LBV Wd1-243 \citep{Ritchie2009}, and the binary-interaction product Wd1-5 \citep{Clark2014}.
For the YSG stars, Fig.~\ref{fig:HR_combined} also includes the luminosity estimated by \citet{Clark2020}, who actually consider them YHG, as an open square.
All luminosities have been scaled to the adopted distance of $4.23^{+0.23}_{-0.21}$~kpc.
Following the alignment of the luminosities on a uniform distance scale, a discernible pattern becomes apparent. 
The majority of evolved stars are consistent with the $\sim$5–6\,Myr~age range. 
Conversely, the blue stragglers are situated well above the 5–6 Myr turn-off, which is consistent with rejuvenation through binary mass transfer or merger.
As shown in Fig.~\ref{fig:HR_combined}, a few RSGs published by \citet{Beasor2021, Beasor2023} fall along the $\sim$10~Myr isochrone when their luminosities are rescaled to $4.23^{+0.23}_{-0.21}$~kpc, while the brightest RSGs could still be reconciled with a younger 6.3~Myr solution.
We note that the location of the cool supergiants (RSGs and YSG/YHGs) in the HRD is very sensitive to the method used to estimate their luminosity, and to the adopted individual extinction correction as differential reddening is significant in Wd1. 
This is further discussed in Sect.~\ref{sec:YSGs}.

Beyond the spectro-photometric results discussed above, independent dynamical constraints within Wd1 further support a young, coeval solution. 
The systems W13, WR-B, and W44/L require primary masses of order $\sim$40\,M$_\odot$ \citep{Ritchie2010}, consistent with the $\sim$5–6\,Myr main-sequence turn-off. 

Our detection of a well populated main sequence provides direct evidence that addresses the long-standing age debate. 
The sample of late-O and early-B type stars reported in this work (Fig.~\ref{fig:HR_combined}) defines a clear turn-off at $\sim$5–6.3~Myr, with no indication of the older, fainter turn-off that would correspond to a 10~Myr population. 
For example, we do not observe a population of B-type supergiants at the luminosities and temperatures expected for such ages. 
This is consistent with the fact that no main sequence star hotter than $\sim$26{,}000~K would be expected at 10~Myr.
Even allowing for uncertainties in distance or extinction, the observed stars are too hot and too luminous to be compatible with a 10~Myr solution.

Our findings also align with other recent studies that have pointed back towards a younger age. 
\citet{Negueruela2022} derived a distance of $4.23^{+0.23}_{-0.21}$~kpc and an age of $\sim$5–6~Myr for Wd1, consistent with our results. 
We also note that \citet{Beasor2021} reported a population of pre-main-sequence stars with an estimated age of $\sim$7~Myr, which is broadly consistent with our derived age within the uncertainties. 
By reaching the main sequence, our work shows that the age inferred from the hot stars and the age inferred from the more luminous stars can be reconciled. 
The combined HRD (Fig.~\ref{fig:HR_combined}) is consistent with a single $\sim$5.5~Myr burst of star formation.
We find no need for an older ($\sim$10~Myr) population in Wd1. 

\subsection{YSGs/YHGs and RSGs luminosities}
\label{sec:YSGs}

The YSG/YHG and RSGs population in Wd1 has been central to the ongoing discussion regarding the cluster’s age. 
\citet{Beasor2021,Beasor2023} derived lower luminosities for these stars (by $\sim$0.5~dex) than those originally obtained by \citet{Clark2005}, leading them to propose that the cool supergiants trace an older ($\sim$10~Myr) population within the cluster.

These differences primarily arise from the adopted methodology. 
\citet{Clark2005} relied on the empirical correlation between the O\,\textsc{i}\,7774\,\AA\ line equivalent width and the absolute visual magnitude ($M_V$), and adopted $BC_V=0$, a hypothesis supported by classical bolometric correction scales for A–F supergiants with similar spectral types (e.g. \citet{Humphreys1984}; \citet{Flower1996}).
Instead, \citet{Beasor2021} derived the stellar luminosities by integrating the dereddened SED.
They reconstructed the SED by joining observed near- and mid-infrared photometry with estimated optical and NIR fluxes inferred from intrinsic colour calibrations, and used a black body function to account for the flux blueward of the $U$ band. 
While this procedure may work for RSGs, it is not suited for warmer YSGs/YHGs ($T_{\rm eff}$ $\sim$ 6800–8500~K) as their SED peaks at shorter wavelengths ($\lambda$ $<$ 0.5~µm). 
Further, rather than deriving line of sight extinctions for individual stars, they adopted a fixed \textit{$A_K~= 0.74~\pm~0.08$} value corresponding to the cluster average extinction from \citet{Damineli2016}.
As a consequence, the resulting inferred luminosities will be highly sensitive to how the optical flux is reconstructed and to the adopted extinction.
The observed large range of NIR colours ($1.8~\lesssim~(J-K)_{\rm obs}\lesssim~3.5$) displayed by the cool supergiants in Wd1 \citep{Navarete2022} is indicative of both high and significant differential reddening across the cluster.

We thus re-evaluated the luminosities of the YSG/YHG and RSG stars using a homogeneous, NIR based approach. 
We made use of available $JHK_s$ photometry \citep{Navarete2022}.
For each star, we used the spectral types from \citet{Clark2020}, and \citet{Navarete2022} and we adopted the $T_{\rm eff}$ from either spectral type vs $T_{\rm eff}$ calibrations in the literature or direct $T_{\rm eff}$ determinations when available (\citet{Levesque2005}, \citet{Arevalo2018}, \citet{Navarete2022}). 
Bolometric corrections and intrinsic colours corresponding to the adopted $T_{\rm eff}$ were then obtained from the YBC database \citep{Chen2019}
\footnote{The YBC database \citep{Chen2019} provides bolometric corrections for a wide range of photometric systems.}.
We estimated the extinction for each star from its observed NIR colour excesses, $E(J-K_s)$, $E(J-H)$ and $E(H-K_s)$, adopting the extinction ratios from \citet{Hosek2018}. 
Because results can differ due to photometric uncertainties, we adopted, for each star, the mean values of $A_J$, $A_H$ and $A_{K_s}$ derived with the available colour combinations.
Finally, we computed bolometric magnitudes in each available band, converted them into luminosities, and adopted the mean value as our final luminosity estimate.
We assumed the cluster distance of $4.23^{+0.23}_{-0.21}$~kpc.
The derived extinction towards individual objects, and the adopted effective temperatures and the luminosity used in Fig.~\ref{fig:HR_combined} are provided in Table~\ref{tab:cool_supergiants}.
We verified that adopting \citet{Damineli2016}'s extinction law yields luminosities consistently larger by $\sim$0.04 dex.
With this method, the four RSGs span $\log(L/L_\odot)\simeq 5.0$--5.4, in excellent agreement with the values derived by \citet{Navarete2022}.
Their location in the HRD is consistent with our derived age for Wd1 and no longer requiring an older ($\sim$10~Myr) solution.
For the YSG/YHGs our method yields higher luminosities than those reported by \citet{Beasor2021}, placing these stars between the 5 and 6.3\,Myr isochrones in Fig.~\ref{fig:HR_combined}.
This supports a young, coeval population consistent with the main sequence turn-off age inferred in this work.

\begin{table}
\centering
\small
\caption{Adopted parameters and bolometric luminosities for the cool supergiants.}
\label{tab:cool_supergiants}
\setlength{\tabcolsep}{3.5pt}

\begin{tabular}{lccccccl}
\hline
 &  &  & \multicolumn{2}{c}{Hosek} & \multicolumn{2}{c}{Damineli} \\
ID & SpT & $T_{\rm eff}$ & $A_{K_s}$ & $\log(L/L_\odot)$ & $A_{K_s}$ & $\log(L/L_\odot)$ \\
 &  & (K) & (mag) & (dex) & (mag) & (dex) \\
\hline
W237  & M3--4\,I     & 3550 & 0.54 & 5.20 & 0.66 & 5.24 \\
W20   & M3--4\,I     & 3550 & 0.76 & 4.98 & 0.93 & 5.04 \\
W75   & M1\,Ia       & 4000 & 1.01 & 5.14 & 1.24 & 5.22 \\
W26   & M1--1.5\,Ia  & 3780 & 0.54 & 5.35 & 0.66 & 5.39 \\
W4    & F3\,Ia$+$    & 6700 & 0.49 & 5.25 & 0.61 & 5.29 \\
W8a   & F8\,Ia$+$    & 6200 & 1.26 & 5.47 & 1.55 & 5.56 \\
W32   & F5\,Ia$+$    & 6500 & 0.61 & 5.45 & 0.74 & 5.49 \\
W16a  & A5\,Ia$+$    & 8200 & 0.63 & 5.40 & 0.77 & 5.45 \\
W12a  & F1\,Ia$+$    & 7150 & 0.66 & 5.22 & 0.81 & 5.27 \\
W265  & F1--5\,Ia$+$ & 6750 & 0.69 & 5.18 & 0.69 & 5.23 \\
\hline
\end{tabular}
\tablefoot{Extinction and luminosities are derived using the extinction laws of \citet{Hosek2018} and \citet{Damineli2016}. 
For reference, \citet{Beasor2021} adopted a cluster average extinction of $A_K = 0.74 \pm 0.08$ for Wd1.}
\end{table}

The observed YSG to RSG ratio in Wd1 is inconsistent with single star evolutionary models, which predict more RSGs than YSGs given the longer duration of the RSG phase.
This discrepancy is further driven by the few YSGs produced by the models.
In addition, the exact number of YSG/YHGs is sensitive to the definition of these objects (by some standards, W16a could be counted as a BSG).
Finally, at the young age derived here ($\sim$5.5~Myr), the progenitors of these stars have initial masses $\geq$30~$M_{\odot}$, for which the RSG phase is very short and model predictions are highly uncertain \citep{Ekstrom2012}.
Instead, we suggest that the population of YSGs and RSGs in Wd1 is heavily influenced by binary physics, and that their relative distribution is not a good indicator of the cluster age.
Indeed, \citet{Wang2025} demonstrated that stellar mergers can drastically alter the luminosity distribution of RSGs, while the YHG HR~5171A has been shown to be an interacting binary \citep{Chesneau2014}.
Regarding the RSGs in Wd1, the ionized nebula surrounding W26 \citep{Wright2014} and the resolved circumstellar shells detected by JWST/MIRI around several cool supergiants \citep{Guarcello2025} may be due to episodic high mass-loss, but may also point to binary interaction.
We therefore do not attempt to reconcile the YSG to RSG ratio with single-star evolutionary models, and instead note that the HRD positions of all cool supergiants are consistent with the $\sim$5--6.3~Myr age anchored by the main-sequence turn-off.

\subsection{Be stars}

\begin{figure*}[t]   % two-column figure
  \centering
  \includegraphics[width=\textwidth]{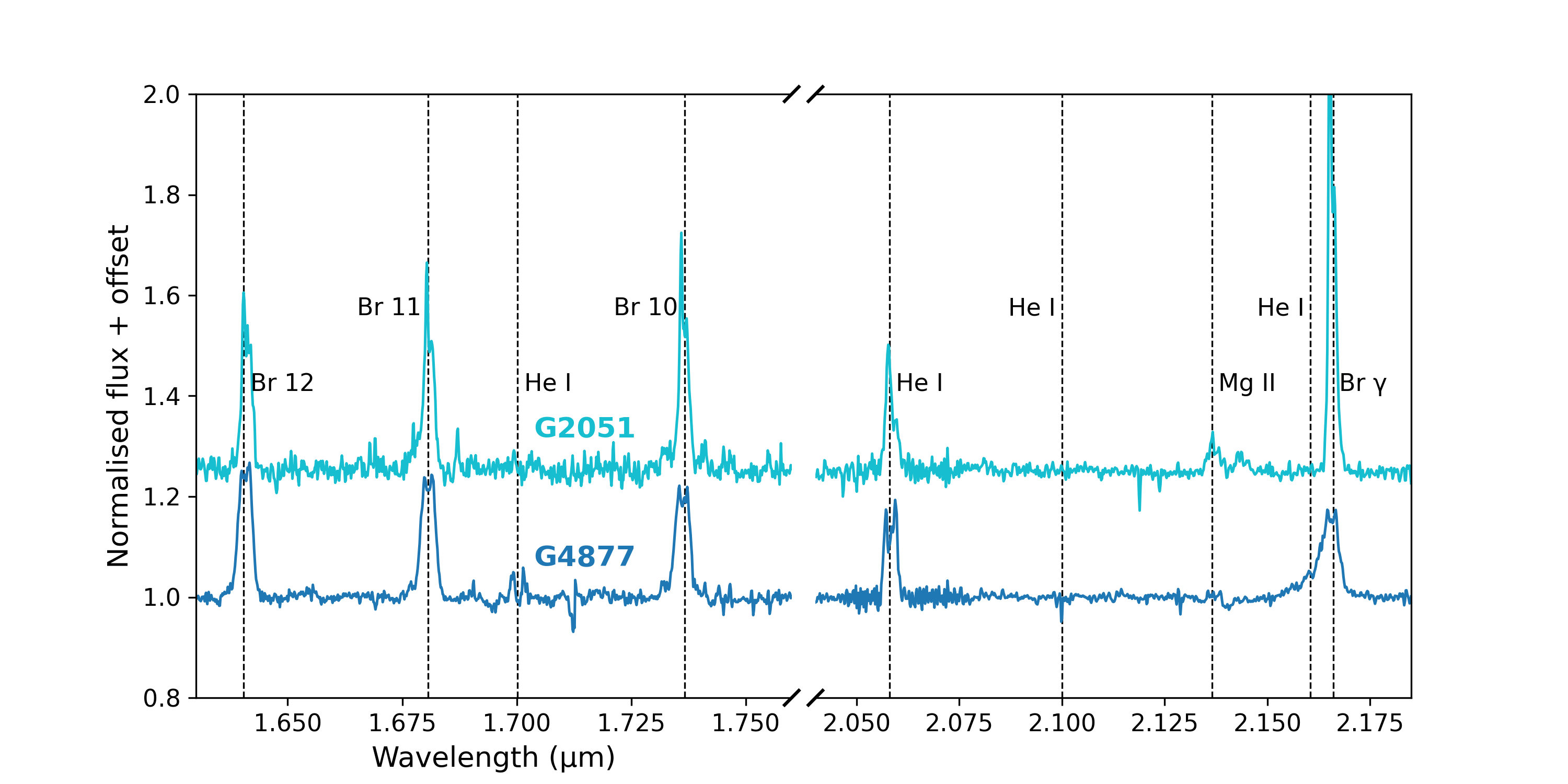}
  \caption{Normalised $H$- and $K$-band spectra of the two Be stars identified in our sample, offset vertically for clarity. 
  Object IDs are labelled above the $H$-band spectra.
  The spectra exhibit the hydrogen emission lines (Br\,$\gamma$ and Br\,10–12) and He\,\textsc{i} and Mg\,\textsc{ii} transitions, typical of Be stars.}
  \label{fig:JH_Be_Star}
\end{figure*}

An interesting outcome of this spectroscopic survey is the detection of two objects with emission lines, namely G2051 and G4877 (see Fig.~\ref{fig:JH_Be_Star}). 
G2051 and G4877 can be identified with Gaia DR3 5940199912246646656 and 5940106007092363648, respectively. 
Both objects have astrometric data compatible (within very large errors) with cluster membership, although G2051 has (BP-RP)=3.8, somewhat bluer than the majority of members. 

There is a third Be star reported in the field, W1004 \citep{Clark2020}. 
This object was cross-identified with Gaia DR3 5940106281970191104, a $G=18.5$ star with no astrometric or colour information, by \citet{Negueruela2022}. 
However, it is blended with Gaia DR3 5940106277666874752, a much brighter ($G=16.3$) star only $0.\!\!^{\prime\prime}7$ away. 
The Gaia astrometric and photometric data are inconclusive due to the high RUWE ($\approx3.7$) and do not allow us to confirm cluster membership.
Although \citet{Clark2020} noted that its interstellar features are typical of Wd1 members, a foreground nature cannot be excluded.
We therefore do not include W1004 in the following discussion.

Be stars are quite numerous in massive clusters of older ages \citep[e.g.,][]{McSwain2005}.
There is little consensus about the cluster age at which the Be fraction is higher, but the number of Be stars around the turn-off can be very high in clusters with ages around 15~Myr, such as NGC~7419 \citep{Marco2013}. 
Significant numbers of Be stars around the turn-off are found in NGC~6871, which is about 10~Myr or slightly younger \citep{Negueruela2004}.
The presence of Be stars decreases in clusters of ages similar to Wd1, such as M~29 (NGC~6913) or NGC~6231 \citep{Negueruela2004,McSwain2005}.
The detection of only two Be stars in our whole sample seems, therefore, consistent with our age estimate for the cluster.
The Be stars in NGC~6913, NGC~6231 and Wd1 are the youngest Be stars known, and their position within the HRD of their respective clusters may shed light on their formation mechanism. 
Unfortunately, we cannot derive astrophysical parameters from the observations presented here, as emission lines contaminate many of the diagnostic lines.

\section{Conclusions}\label{sec:conclusions}

The NIR spectroscopic detection of the main sequence in Wd1 closes the long-standing uncertainty caused by the cluster’s heavy optical extinction. 
We identify $\sim$50 faint O9–B1 III-V members, locate the turn-off at \(T_{\!\rm eff}\simeq30,000\)~K and \(L_{\ast}\simeq1\times10^{5}\,L_\odot\), and show that enriched metallicity Geneva isochrones with ages between 5 and 6.3~Myr reproduce the entire HRD  when scaled to $4.23^{+0.23}_{-0.21}$~kpc with the \citet{Hosek2018} extinction law, yielding an age of $5.5\pm1.0$~Myr.

Wd1 emerges as a coeval, young, massive system in agreement with the $\sim$5--6\,Myr ages previously inferred from its most massive evolved stars, and now confirmed by direct observation of its previously hidden unevolved members.
Within our sensitivity and resolution limits, we detect no stellar population consistent with a $\sim$10~Myr age or extended star formation history.
Adopting homogeneous, NIR based luminosity estimates for the cool supergiants, with line of sight extinctions and bolometric corrections appropriate for their spectral types, brings both the YSG/YHG and RSG populations into agreement with a $\sim$5--6.3\,Myr cluster age.
This supports a single, coeval burst of star formation in Wd1, without the need to invoke an older stellar population.

The HRD solution remains stable against reasonable shifts in distance, metallicity, and extinction curve.
None of these variations alters the derived age by more than $\sim$1~Myr (see Appendix.~\ref{sec:Appendix_parameters}). 
The tight $\sim$5.5~Myr age constraint provided here offers a crucial anchor for ongoing large-scale programmes \citep[e.g.][]{Guarcello2024} that aim to characterise Wd1 across its entire mass spectrum.

Finally, our multi-epoch spectroscopic observations reveal a remarkably high incidence of radial-velocity variability among the unevolved members of Wd1. 
We identify $\sim$$65\%$ candidate spectroscopic binaries among the OB stars with multi-epoch coverage, a fraction larger than previously reported by the optical FLAMES data \citep{Ritchie2022}. 
The high incidence of radial-velocity variability among main-sequence stars, together with the presence of evolved objects, underscores Wd1’s value as a natural laboratory for studying binary interaction in massive star evolution. 

\begin{acknowledgements}
We thank our anonymous referee for careful and constructive comments.
R.Castellanos gratefully  acknowledges the financial support through the PRE2020-096167 grant provided by the Spanish MCIN.
R. Castellanos, F. Najarro, M. Garcia, and L.R. Patrick gratefully acknowledge support by grant PID2022-137779OB-C41, and M. Garcia and L.R. Patrick further acknowledge grant PID2022-140483NB-C22, funded by the Spanish Ministry of Science, Innovation and Universities/State Agency of Research
MICIU/AEI/10.13039/501100011033 and by “ERDF A way of making Europe”. 
F. Najarro and M. Garcia also acknowledge grant MAD4SPACE, TEC-2024/TEC-182 from Comunidad de Madrid (Spain). 
T. Shenar acknowledges support from the Israel Science Foundation (ISF) under grant number 0603225041 and from the European Research Council (ERC) under the European Union's Horizon 2020 research and innovation program (grant agreement 101164755/METAL).
I. Negueruela is partially supported by the MCIU and AEI under grants PID2021-122397NB-C22 and PID2024-159329NB-C22.
M. G. Guarcello acknowledges the INAF grant 1.05.12.05.03.
Based on observations collected at the European Southern Observatory under ESO programmes 109.233D.001 and data obtained from the ESO Science Archive Facility with DOI: https://doi.org/10.18727/archive/71.
\end{acknowledgements}

\section*{Data availability}
VLT data are available in the ESO Science Archive Facility (\url{http://archive.eso.org/wdb/wdb/adp/phase3_main/form}).

% Bibliografía
\bibliographystyle{aa}
\bibliography{Bibliography/references}

\begin{thebibliography}{56}
\expandafter\ifx\csname natexlab\endcsname\relax\def\natexlab#1{#1}\fi

\bibitem[{{Aghakhanloo} {et~al.}(2021){Aghakhanloo}, {Murphy}, {Smith}, {Parejko}, {D{\'\i}az-Rodr{\'\i}guez}, {Drout}, {Groh}, {Guzman}, \& {Stassun}}]{Aghakhanloo2021}
{Aghakhanloo}, M., {Murphy}, J.~W., {Smith}, N., {et~al.} 2021, Research Notes of the American Astronomical Society, 5, 14

\bibitem[{{Anastasopoulou} {et~al.}(2024){Anastasopoulou}, {Guarcello}, {Flaccomio}, {Sciortino}, {Benatti}, {De Becker}, {Wright}, {Drake}, {Albacete-Colombo}, {Andersen}, {Argiroffi}, {Bayo}, {Castellanos}, {Gennaro}, {Grebel}, {Miceli}, {Najarro}, {Negueruela}, {Prisinzano}, {Ritchie}, {Robberto}, {Sabbi}, \& {Zeidler}}]{Anastasopoulou2024}
{Anastasopoulou}, K., {Guarcello}, M.~G., {Flaccomio}, E., {et~al.} 2024, \aap, 690, A25

\bibitem[{{Andersen} {et~al.}(2017){Andersen}, {Gennaro}, {Brandner}, {Stolte}, {de Marchi}, {Meyer}, \& {Zinnecker}}]{Andersen2017}
{Andersen}, M., {Gennaro}, M., {Brandner}, W., {et~al.} 2017, \aap, 602, A22

\bibitem[{Ar{\'e}valo(2018)}]{Arevalo2018}
Ar{\'e}valo, A. d. L. E.~R. 2018, Master's thesis, Instituto de Astronomia, Geof{\'i}sica e Ci{\^e}ncias Atmosf{\'e}ricas, University of S{\~a}o Paulo, doi:10.11606/D.14.2019.tde-12092018-161841

\bibitem[{Beasor {et~al.}(2021)Beasor, Davies, Smith, Gehrz, \& Figer}]{Beasor2021}
Beasor, E.~R., Davies, B., Smith, N., Gehrz, R.~D., \& Figer, D.~F. 2021, Astrophysical Journal, 912, 16

\bibitem[{{Beasor} {et~al.}(2023){Beasor}, {Smith}, \& {Andrews}}]{Beasor2023}
{Beasor}, E.~R., {Smith}, N., \& {Andrews}, J.~E. 2023, \apj, 952, 113

\bibitem[{{Bodensteiner} {et~al.}(2021){Bodensteiner}, {Sana}, {Wang}, {Langer}, {Mahy}, {Banyard}, {de Koter}, {de Mink}, {Evans}, {G{\"o}tberg}, {Patrick}, {Schneider}, \& {Tramper}}]{Bodensteiner2021}
{Bodensteiner}, J., {Sana}, H., {Wang}, C., {et~al.} 2021, \aap, 652, A70

\bibitem[{{Brandner} {et~al.}(2008){Brandner}, {Clark}, {Stolte}, {Waters}, {Negueruela}, \& {Goodwin}}]{Brandner2008}
{Brandner}, W., {Clark}, J.~S., {Stolte}, A., {et~al.} 2008, \aap, 478, 137

\bibitem[{{Cardelli} {et~al.}(1989){Cardelli}, {Clayton}, \& {Mathis}}]{Cardelli1989}
{Cardelli}, J.~A., {Clayton}, G.~C., \& {Mathis}, J.~S. 1989, \apj, 345, 245

\bibitem[{{Chen} {et~al.}(2019){Chen}, {Girardi}, {Fu}, {Bressan}, {Aringer}, {Dal Tio}, {Pastorelli}, {Marigo}, {Costa}, \& {Zhang}}]{Chen2019}
{Chen}, Y., {Girardi}, L., {Fu}, X., {et~al.} 2019, \aap, 632, A105

\bibitem[{{Chesneau} {et~al.}(2014){Chesneau}, {Meilland}, {Chapellier}, {Millour}, {van Genderen}, {Naz{\'e}}, {Smith}, {Spang}, {Smoker}, {Dessart}, {Kanaan}, {Bendjoya}, {Feast}, {Groh}, {Lobel}, {Nardetto}, {Otero}, {Oudmaijer}, {Tekola}, {Whitelock}, {Arcos}, {Cur{\'e}}, \& {Vanzi}}]{Chesneau2014}
{Chesneau}, O., {Meilland}, A., {Chapellier}, E., {et~al.} 2014, \aap, 563, A71

\bibitem[{{Clark} {et~al.}(2019){Clark}, {Najarro}, {Negueruela}, {Ritchie}, {Gonz{\'a}lez-Fern{\'a}ndez}, \& {Lohr}}]{Clark2019}
{Clark}, J.~S., {Najarro}, F., {Negueruela}, I., {et~al.} 2019, \aap, 623, A83

\bibitem[{Clark {et~al.}(2005)Clark, Negueruela, Crowther, \& Goodwin}]{Clark2005}
Clark, J.~S., Negueruela, I., Crowther, P.~A., \& Goodwin, S.~P. 2005, Astronomy \& Astrophysics, 434, 949

\bibitem[{{Clark} {et~al.}(2014){Clark}, {Ritchie}, {Najarro}, {Langer}, \& {Negueruela}}]{Clark2014}
{Clark}, J.~S., {Ritchie}, B.~W., {Najarro}, F., {Langer}, N., \& {Negueruela}, I. 2014, \aap, 565, A90

\bibitem[{{Clark} {et~al.}(2020){Clark}, {Ritchie}, \& {Negueruela}}]{Clark2020}
{Clark}, J.~S., {Ritchie}, B.~W., \& {Negueruela}, I. 2020, \aap, 635, A187

\bibitem[{{Clark} {et~al.}(2011){Clark}, {Ritchie}, {Negueruela}, {Crowther}, {Damineli}, {Jablonski}, \& {Langer}}]{Clark2011}
{Clark}, J.~S., {Ritchie}, B.~W., {Negueruela}, I., {et~al.} 2011, \aap, 531, A28

\bibitem[{Crowther {et~al.}(2006)Crowther, Hadfield, Clark, Negueruela, \& Vacca}]{Crowther2006}
Crowther, P.~A., Hadfield, L.~J., Clark, J.~S., Negueruela, I., \& Vacca, W.~D. 2006, Monthly Notices of the RAS, 372, 1407

\bibitem[{Damineli {et~al.}(2016)Damineli, Almeida, Blum, Damineli, Navarete, Rubinho, \& Teodoro}]{Damineli2016}
Damineli, A., Almeida, L.~A., Blum, R.~D., {et~al.} 2016, Monthly Notices of the RAS, 463, 2653

\bibitem[{{Davies} \& {Beasor}(2019)}]{DaviesBeasor2019}
{Davies}, B. \& {Beasor}, E.~R. 2019, \mnras, 486, L10

\bibitem[{{Davies} {et~al.}(2013){Davies}, {Agudo Berbel}, {Wiezorrek}, {Cirasuolo}, {F{\"o}rster Schreiber}, {Jung}, {Muschielok}, {Ott}, {Ramsay}, {Schlichter}, {Sharples}, \& {Wegner}}]{Davies2013}
{Davies}, R.~I., {Agudo Berbel}, A., {Wiezorrek}, E., {et~al.} 2013, \aap, 558, A56

\bibitem[{{Ekstr{\"o}m} {et~al.}(2012){Ekstr{\"o}m}, {Georgy}, {Eggenberger}, {Meynet}, {Mowlavi}, {Wyttenbach}, {Granada}, {Decressin}, {Hirschi}, {Frischknecht}, {Charbonnel}, \& {Maeder}}]{Ekstrom2012}
{Ekstr{\"o}m}, S., {Georgy}, C., {Eggenberger}, P., {et~al.} 2012, \aap, 537, A146

\bibitem[{{Fenech} {et~al.}(2018){Fenech}, {Clark}, {Prinja}, {Dougherty}, {Najarro}, {Negueruela}, {Richards}, {Ritchie}, \& {Andrews}}]{Fenech2018}
{Fenech}, D.~M., {Clark}, J.~S., {Prinja}, R.~K., {et~al.} 2018, \aap, 617, A137

\bibitem[{{Flower}(1996)}]{Flower1996}
{Flower}, P.~J. 1996, \apj, 469, 355

\bibitem[{{Freudling} {et~al.}(2013){Freudling}, {Romaniello}, {Bramich}, {Ballester}, {Forchi}, {Garc{\'\i}a-Dabl{\'o}}, {Moehler}, \& {Neeser}}]{Freudling2013}
{Freudling}, W., {Romaniello}, M., {Bramich}, D.~M., {et~al.} 2013, \aap, 559, A96

\bibitem[{{Gaia Collaboration} {et~al.}(2021){Gaia Collaboration}, Brown, Vallenari, Prusti, de~Bruijne, \& Babusiaux}]{GaiaEDR3}
{Gaia Collaboration}, Brown, A. G.~A., Vallenari, A., {et~al.} 2021, Astronomy \& Astrophysics, 649, A1

\bibitem[{{Gennaro} {et~al.}(2011){Gennaro}, {Brandner}, {Stolte}, \& {Henning}}]{Gennaro2011}
{Gennaro}, M., {Brandner}, W., {Stolte}, A., \& {Henning}, T. 2011, \mnras, 412, 2469

\bibitem[{{Guarcello} {et~al.}(2025{\natexlab{a}}){Guarcello}, {Almendros-Abad}, {Lovell}, {Monsch}, {Mu{\v{z}}i{\'c}}, {Mart{\'\i}nez-Galarza}, {Drake}, {Anastasopoulou}, {Andersen}, {Argiroffi}, {Bayo}, {Bonito}, {Capela}, {Damiani}, {Gennaro}, {Ginsburg}, {Grebel}, {Hora}, {Moraux}, {Najarro}, {Negueruela}, {Prisinzano}, {Richardson}, {Ritchie}, {Robberto}, {Rom}, {Sabbi}, {Sciortino}, {Umana}, {Winter}, {Wright}, \& {Zeidler}}]{Guarcello2024}
{Guarcello}, M.~G., {Almendros-Abad}, V., {Lovell}, J.~B., {et~al.} 2025{\natexlab{a}}, \aap, 693, A120

\bibitem[{{Guarcello} {et~al.}(2025{\natexlab{b}}){Guarcello}, {Almendros-Abad}, {Lovell}, {Monsch}, {Mu{\v{z}}i{\'c}}, {Mart{\'\i}nez-Galarza}, {Drake}, {Anastasopoulou}, {Andersen}, {Argiroffi}, {Bayo}, {Bonito}, {Capela}, {Damiani}, {Gennaro}, {Ginsburg}, {Grebel}, {Hora}, {Moraux}, {Najarro}, {Negueruela}, {Prisinzano}, {Richardson}, {Ritchie}, {Robberto}, {Rom}, {Sabbi}, {Sciortino}, {Umana}, {Winter}, {Wright}, \& {Zeidler}}]{Guarcello2025}
{Guarcello}, M.~G., {Almendros-Abad}, V., {Lovell}, J.~B., {et~al.} 2025{\natexlab{b}}, \aap, 693, A120

\bibitem[{{Guarcello} {et~al.}(2024){Guarcello}, {Flaccomio}, {Albacete-Colombo}, {Almendros-Abad}, {Anastasopoulou}, {Andersen}, {Argiroffi}, {Bayo}, {Bartlett}, {Bastian}, {De Becker}, {Best}, {Bonito}, {Borghese}, {Calzetti}, {Castellanos}, {Cecchi-Pestellini}, {Clark}, {Clarke}, {Coti Zelati}, {Damiani}, {Drake}, {Gennaro}, {Ginsburg}, {Grebel}, {Hora}, {Israel}, {Lawrence}, {Locci}, {Mapelli}, {Martinez-Galarza}, {Micela}, {Miceli}, {Moraux}, {Muzic}, {Najarro}, {Negueruela}, {Nota}, {Pallanca}, {Prisinzano}, {Ritchie}, {Robberto}, {Rom}, {Sabbi}, {Scholz}, {Sciortino}, {Trigilio}, {Umana}, {Winter}, {Wright}, \& {Zeidler}}]{Guarcello2024I}
{Guarcello}, M.~G., {Flaccomio}, E., {Albacete-Colombo}, J.~F., {et~al.} 2024, \aap, 682, A49

\bibitem[{{Hanson} {et~al.}(2005){Hanson}, {Kudritzki}, {Kenworthy}, {Puls}, \& {Tokunaga}}]{Hanson2005}
{Hanson}, M.~M., {Kudritzki}, R.~P., {Kenworthy}, M.~A., {Puls}, J., \& {Tokunaga}, A.~T. 2005, \apjs, 161, 154

\bibitem[{{Hillier} \& {Miller}(1998)}]{HillierMiller1998}
{Hillier}, D.~J. \& {Miller}, D.~L. 1998, \apj, 496, 407

\bibitem[{{Hosek} {et~al.}(2018){Hosek}, {Lu}, {Anderson}, {Do}, {Schlafly}, {Ghez}, {Clarkson}, {Morris}, \& {Albers}}]{Hosek2018}
{Hosek}, Jr., M.~W., {Lu}, J.~R., {Anderson}, J., {et~al.} 2018, \apj, 855, 13

\bibitem[{{Howarth}(1983)}]{Howarth1983}
{Howarth}, I.~D. 1983, \mnras, 203, 301

\bibitem[{{Humphreys} \& {McElroy}(1984)}]{Humphreys1984}
{Humphreys}, R.~M. \& {McElroy}, D.~B. 1984, \apj, 284, 565

\bibitem[{{Koumpia} \& {Bonanos}(2012)}]{KoumpiaBonanos2012}
{Koumpia}, E. \& {Bonanos}, A.~Z. 2012, \aap, 547, A30

\bibitem[{{Levesque} {et~al.}(2005){Levesque}, {Massey}, {Olsen}, {Plez}, {Maeder}, \& {Meynet}}]{Levesque2005}
{Levesque}, E.~M., {Massey}, P., {Olsen}, K.~A.~G., {et~al.} 2005, in American Astronomical Society Meeting Abstracts, Vol. 207, American Astronomical Society Meeting Abstracts, 182.13

\bibitem[{{Lim} {et~al.}(2013){Lim}, {Chun}, {Sung}, {Park}, {Lee}, {Sohn}, {Hur}, \& {Bessell}}]{Lim2013}
{Lim}, B., {Chun}, M.-Y., {Sung}, H., {et~al.} 2013, \aj, 145, 46

\bibitem[{{Marco} \& {Negueruela}(2013)}]{Marco2013}
{Marco}, A. \& {Negueruela}, I. 2013, \aap, 552, A92

\bibitem[{{McSwain} \& {Gies}(2005)}]{McSwain2005}
{McSwain}, M.~V. \& {Gies}, D.~R. 2005, \apj, 622, 1052

\bibitem[{{Muno} {et~al.}(2006){Muno}, {Clark}, {Crowther}, {Dougherty}, {de Grijs}, {Law}, {McMillan}, {Morris}, {Negueruela}, {Pooley}, {Portegies Zwart}, \& {Yusef-Zadeh}}]{Muno2006}
{Muno}, M.~P., {Clark}, J.~S., {Crowther}, P.~A., {et~al.} 2006, \apjl, 636, L41

\bibitem[{{Navarete} {et~al.}(2022){Navarete}, {Damineli}, {Ramirez}, {Rocha}, \& {Almeida}}]{Navarete2022}
{Navarete}, F., {Damineli}, A., {Ramirez}, A.~E., {Rocha}, D.~F., \& {Almeida}, L.~A. 2022, \mnras, 516, 1289

\bibitem[{{Negueruela}(2004)}]{Negueruela2004}
{Negueruela}, I. 2004, Astronomische Nachrichten, 325, 380

\bibitem[{{Negueruela} {et~al.}(2022){Negueruela}, {Alfaro}, {Dorda}, {Marco}, {Ma{\'\i}z Apell{\'a}niz}, \& {Gonz{\'a}lez-Fern{\'a}ndez}}]{Negueruela2022}
{Negueruela}, I., {Alfaro}, E.~J., {Dorda}, R., {et~al.} 2022, \aap, 664, A146

\bibitem[{Negueruela {et~al.}(2010)Negueruela, Clark, \& Ritchie}]{Negueruela2010}
Negueruela, I., Clark, J.~S., \& Ritchie, B.~W. 2010, Astronomy \& Astrophysics, 516, A78

\bibitem[{{Negueruela} {et~al.}(2024){Negueruela}, {Sim{\'o}n-D{\'\i}az}, {de Burgos}, {Casasbuenas}, \& {Beck}}]{Negueruela2024}
{Negueruela}, I., {Sim{\'o}n-D{\'\i}az}, S., {de Burgos}, A., {Casasbuenas}, A., \& {Beck}, P.~G. 2024, \aap, 690, A176

\bibitem[{{Ritchie} {et~al.}(2010){Ritchie}, {Clark}, {Negueruela}, \& {Langer}}]{Ritchie2010}
{Ritchie}, B.~W., {Clark}, J.~S., {Negueruela}, I., \& {Langer}, N. 2010, \aap, 520, A48

\bibitem[{{Ritchie} {et~al.}(2009){Ritchie}, {Clark}, {Negueruela}, \& {Najarro}}]{Ritchie2009}
{Ritchie}, B.~W., {Clark}, J.~S., {Negueruela}, I., \& {Najarro}, F. 2009, \aap, 507, 1597

\bibitem[{{Ritchie} {et~al.}(2022){Ritchie}, {Clark}, {Negueruela}, \& {Najarro}}]{Ritchie2022}
{Ritchie}, B.~W., {Clark}, J.~S., {Negueruela}, I., \& {Najarro}, F. 2022, \aap, 660, A89

\bibitem[{Rosslowe(2015)}]{Rosslowe2015}
Rosslowe, C. 2015, PhD thesis, University of Sheffield

\bibitem[{{Sana} {et~al.}(2013){Sana}, {de Koter}, {de Mink}, {Dunstall}, {Evans}, {H{\'e}nault-Brunet}, {Ma{\'\i}z Apell{\'a}niz}, {Ram{\'\i}rez-Agudelo}, {Taylor}, {Walborn}, {Clark}, {Crowther}, {Herrero}, {Gieles}, {Langer}, {Lennon}, \& {Vink}}]{Sana2013}
{Sana}, H., {de Koter}, A., {de Mink}, S.~E., {et~al.} 2013, \aap, 550, A107

\bibitem[{{Sharples} {et~al.}(2013){Sharples}, {Bender}, {Agudo Berbel}, {Bezawada}, {Castillo}, {Cirasuolo}, {Davidson}, {Davies}, {Dubbeldam}, {Fairley}, {Finger}, {F{\"o}rster Schreiber}, {Gonte}, {Hess}, {Jung}, {Lewis}, {Lizon}, {Muschielok}, {Pasquini}, {Pirard}, {Popovic}, {Ramsay}, {Rees}, {Richter}, {Riquelme}, {Rodrigues}, {Saviane}, {Schlichter}, {Schmidtobreick}, {Segovia}, {Smette}, {Szeifert}, {van Kesteren}, {Wegner}, \& {Wiezorrek}}]{Sharples2013}
{Sharples}, R., {Bender}, R., {Agudo Berbel}, A., {et~al.} 2013, The Messenger, 151, 21

\bibitem[{Smette {et~al.}(2015)Smette, Sana, Noll, Horst, Kausch, Kimeswenger, Barden, Szyszka, Jones, Gallenne, {et~al.}}]{Smette2015}
Smette, A., Sana, H., Noll, S., {et~al.} 2015, Astronomy \& Astrophysics, 576, A77

\bibitem[{{Wang} {et~al.}(2025){Wang}, {Patrick}, {Schootemeijer}, {de Mink}, {Langer}, {Britavskiy}, {Xu}, {Bodensteiner}, {Laplace}, {Valli}, {Vigna-G{\'o}mez}, {Klencki}, {Justham}, {Johnston}, \& {Ma}}]{Wang2025}
{Wang}, C., {Patrick}, L., {Schootemeijer}, A., {et~al.} 2025, \apjl, 981, L16

\bibitem[{Westerlund(1961)}]{Westerlund1961}
Westerlund, B. 1961, Astron. J., 66, 57

\bibitem[{{Wright} {et~al.}(2014){Wright}, {Wesson}, {Drew}, {Barentsen}, {Barlow}, {Walsh}, {Zijlstra}, {Drake}, {Eisl{\"o}ffel}, \& {Farnhill}}]{Wright2014}
{Wright}, N.~J., {Wesson}, R., {Drew}, J.~E., {et~al.} 2014, \mnras, 437, L1

\bibitem[{{Yusof} {et~al.}(2022){Yusof}, {Hirschi}, {Eggenberger}, {Ekstr{\"o}m}, {Georgy}, {Sibony}, {Crowther}, {Meynet}, {Kassim}, {Harun}, {Maeder}, {Groh}, {Farrell}, \& {Murphy}}]{Yusof2022}
{Yusof}, N., {Hirschi}, R., {Eggenberger}, P., {et~al.} 2022, \mnras, 511, 2814

\end{thebibliography}

% Apéndices
\onecolumn 
\begin{appendix}

\section{Stellar parameters of the analysed sample}
\label{sec:Appendix_table}
\setlength{\tabcolsep}{4.5pt}%
\renewcommand{\arraystretch}{1.1}%
\begin{longtable}{@{\hspace{0pt}}l@{\hspace{6pt}}c@{\hspace{6pt}}c@{\hspace{6pt}}c@{\hspace{6pt}}c@{\hspace{6pt}}c@{\hspace{6pt}}c@{\hspace{6pt}}c}
\caption{Stellar parameters for the analysed sample.}
\label{tab:Stellar_parameters}\\
\toprule
       Star &  RA (J2000) &  Dec (J2000) &  $T_{\rm eff}$ (kK) & $\log L/L_{\odot}$ &  $\log g$ (cgs) &                             Spectral type & Binarity status \\
\midrule
\endfirsthead
\caption[]{(cont.)} \\
\label{tab:Stellar_parameters}\\
\toprule
       Star &  RA (J2000) &  Dec (J2000) &  $T_{\rm eff}$ (kK) & $\log L/L_{\odot}$ &  $\log g$ (cgs) &                             Spectral type & Binarity status \\
\midrule
\endhead
\midrule
\multicolumn{8}{r}{{Continued on next page}} \\
\midrule
\endfoot

\bottomrule
\multicolumn{8}{@{}p{\textwidth}@{}}{\smallskip\textit{Notes.} $T_{\rm eff}$ is given in kK; luminosity as $\log\,L/L_{\odot}$ and surface gravity as $\log\,g$ (cgs). 
Star identification codes are from \citet{Clark2020} unless otherwise noted. 
Spectral types were derived in this work unless marked. 
Column `Binarity status': ($\checkmark$) = candidate identified in this work; ($\blacktriangle$) = binary flagged in \citet{Ritchie2022}; ($\triangle$) = candidate flagged in \citet{Ritchie2022}. 
$^{1}$\,Identification code from \citet{Gennaro2011}. 
$^{2}$\,Previous spectral type from \citet{Clark2020}.
$^{3}$\,OB star confirmed in this work; ID from \citet{GaiaEDR3}.} \\
\endlastfoot
        W2a & 16:46:59.72 & -45:50:51.19 &               22.0 &      5.34 &      2.85 &      B2 Ia$^{2}$ &   $\blacktriangle$ \\
        W6b & 16:47:02.93 & -45:50:22.54 &               28.5 &    4.32 &      3.75 &        B0-0.5 III (O9.5 III$^{2}$) &   $\blacktriangle$ \\
        W10 & 16:47:03.34 & -45:50:34.60 &               29.0 &               5.72 &      3.60 & B0.5 I+OB$^{2}$ &   $\blacktriangle$ \\
        W15 & 16:47:06.63 & -45:50:29.79 &               30.3 &      4.99 &      3.42 &        O9-9.5 II (O9 Ib$^{2}$) &            $\triangle$ \\
        W17 & 16:47:06.25 & -45:50:49.35 &               30.3 &       5.11 &      3.42 &       O9 Iab$^{2}$ &            $\triangle$ \\
        W18 & 16:47:05.70 & -45:50:50.54 &               23.5 &        5.23 &      3.00 &        B0.5 Ia$^{2}$ &                    \\
        W21 & 16:47:01.11 & -45:51:13.64 &     23.5 &               5.28 &      3.00 &        B0.5 Ia$^{2}$ &                    \\
        W24 & 16:47:02.15 & -45:51:12.64 &               30.3 &               5.43 &      3.42 &     O9-9.5 II (O9 Iab$^{2}$) &   $\blacktriangle$ \\
        W29 & 16:47:04.41 & -45:50:39.99 &               30.3 &               5.19 &      3.42 &    O9-9.5 II (O9 Ib$^{2}$) &                    \\
        W37 & 16:47:06.01 & -45:50:47.53 &               30.2 &               5.14 &      3.45 &       O9.5 Ib$^{2}$ &                    \\
        W38 & 16:47:02.86 & -45:50:46.20 &               30.2 &         5.13 &      3.60 &    O9-9.5 I-II (O9 Iab$^{2}$) &                    \\
       W43c & 16:47:03.76 & -45:50:58.49 &               30.3 &               5.16 &      3.42 &    O9-9.5 II (O9 Ib$^{2}$) &                    \\
       W47 &   16:47:02.59 & -45:51:17.91   &  29.2   &  5.05   &  3.60  &    B0-0.5 II-III (O9.5 Iab$^{2}$) &      $\triangle$   \\
       W49 & 16:47:01.90 & -45:50:31.71 &      29.0 &               4.89 &      3.45 &   B0-0.5 II-III (B0 Iab$^{2}$) &                    \\
       W50b & 16:47:01.17 & -45:50:26.87 &     30.2 &               4.63 &      3.75 &      O9.5 III$^{2}$ &                    \\
        W52 & 16:47:01.84 & -45:51:29.43 &         23.5 &               5.54 &      3.00 &        B1.5 Ia$^{2}$ &   $\blacktriangle$ \\
        W55 & 16:46:58.42 & -45:51:31.36 &       28.0 &               5.23 &      3.30 &   B0 Ia$^{2}$ &                    \\
       W56b & 16:46:58.86 & -45:51:46.00 &         30.3 &               4.93 &      3.42 &    O9.5 II-III (O9.5 Ib$^{2}$) &                    \\
        W60 & 16:47:04.13 & -45:51:52.32 &         30.2 &               5.25 &      3.45 &    O9-9.5 I-II (B0 Iab$^{2}$)  &                    \\
       W61b &  16:47:02.56 & -45:51:41.92  &    30.2   &   5.24         &   3.45   &    O9.5 Iab$^{2}$ &                    \\

       W62a & 16:47:02.52 & -45:51:38.09 &        29.5 &               5.05 &      3.60 &     O9-9.5 II-III (B0.5 Ib$^{2}$) &                    \\
        W84 & 16:46:59.05 & -45:50:28.35 &        30.7 &               4.93 &      3.50 &    O9.5 Ib$^{2}$ &                    \\
        W86 & 16:46:57.16 & -45:50:10.00 &       30.3 &               4.88 &      3.42 &         O9.5 Ib$^{2}$ &                    \\
       W232 & 16:47:01.43 & -45:52:35.16 &       30.3 &               5.11 &      3.42 &      O9-9.5 II-III (B0 Iab$^{2}$) &   $\blacktriangle$ \\
       W238 & 16:47:04.42 & -45:52:27.87 &       26.0 &               5.41 &      3.30 &                  B1 Iab$^{2}$ &                    \\
      W1003 & 16:46:52.35 & -45:52:03.41 &       26.4 &               4.17 &      4.00 &  B0-0.5 V &          \\
      W1006 & 16:46:54.48 & -45:53:30.10 &       29.0 &               4.28 &      3.75 &       O9-9.5 III$^{2}$ &    $\triangle$   \\
      W1007 & 16:46:54.94 & -45:50:05.98 &       30.7 &               4.99 &      3.60 &               O9.5 III$^{2}$ &                    \\
      W1008 & 16:46:55.40 & -45:51:54.33 &      30.5 &               4.74 &      3.75 &                  O9.5 II$^{2}$ &                    \\
      W1009 & 16:46:55.91 & -45:51:41.46 &      29.6 &               4.79 &      3.60 &    O9.5 II-III (B0 Ib$^{2}$) &                    \\
      W1014 & 16:46:57.82 & -45:51:19.86 &      28.0 &               4.26 &      3.60 &    B0-0.5 III (O9-9.5 III$^{2}$) &                    \\
      W1015 & 16:46:57.97 & -45:51:40.88 &      30.2 &               4.78 &      3.45 &                  O9 III$^{2}$ &                    \\
      W1017 & 16:46:58.24 & -45:50:34.04 &       31.5 &               4.59 &     4.15 &     O9.5 V (O9-9.5 III$^{2}$) &    $\triangle$    \\
      W1018 &   16:46:58.23 & -45:50:56.99   &  29.5   &   4.71   &  3.75  &   O9-9.5 II-III (O9.5 Iab$^{2}$) &      $\triangle$    \\
      W1019 & 16:46:58.38 & -45:51:49.01 &       28.3 &               4.29 &      3.75 &    B0-0.5 III (O9-9.5 III$^{2}$) &                    \\
      W1022 & 16:46:59.93 & -45:50:25.60 &      30.5 &               4.57 &      3.75 &    O9.5 II$^{2}$ &   $\blacktriangle$ \\
      W1023 & 16:47:00.14 & -45:51:10.51 &       30.2 &               4.96 &      3.60 &         O9 III$^{2}$ &                    \\
      W1026 & 16:47:01.03 & -45:49:48.96 &        29.0 &               4.47 &      3.60 &    B0-0.5 III (O9.5 III$^{2}$) &            $\triangle$ \\
      W1028 & 16:47:01.32 & -45:51:38.24 &        29.0 &               4.39 &      3.75 &  B0-0.5 III (O9-9.5 III$^{2}$) &            $\triangle$ \\
      W1029 & 16:47:01.50 & -45:49:50.27 &       29.4 &               4.62 &      3.60 &      O9.5 III$^{2}$ &        $\triangle$  \\
      W1030 & 16:47:01.68 & -45:52:58.13 &      30.3 &               5.17 &      3.42 &              O9-9.5 II-III (O9.5 Iab$^{2}$) &   $\blacktriangle$ \\
      W1033 & 16:47:02.36 & -45:52:34.41 &       30.7 &               4.93 &      3.50 &    O9-9.5 III (O9-9.5 I-III$^{2}$) &            $\triangle$ \\
      W1034 & 16:47:02.53 & -45:51:48.48 &       31.5 &               4.92 &      3.65 &   O9.5 II-III (O9.5 Iab$^{2}$) &                    \\
      W1036 & 16:47:02.78 & -45:52:12.66 &       30.2 &               4.95 &      3.75 &      O9.5 II-III (O9.5 Iab$^{2}$) &                    \\
      W1043 & 16:47:04.56 & -45:50:59.52 &      30.2 &               4.93 &      3.60 &     O9.5 II-III$^{2}$ &     $\triangle$   \\
      W1044 & 16:47:05.56 & -45:49:51.66 &      29.6 &               4.45 &      3.60 &        B0-0.5 III (O9.5 III$^{2}$) &                    \\
      W1045 &  16:47:05.82 & -45:51:55.08   &    30.2  &   4.71   &   3.45  &   O9.5 II$^{2}$   &                    \\
      W1047 &  16:47:06.11 & -45:52:32.31  &  29.7   &  4.87   &  3.60   &   O9.5 II$^{2}$   &  $\triangle$  \\
      W1048 & 16:47:06.26 & -45:51:03.95 &        28.0 &               5.26 &      3.30 &   B0-0.5 Ia (O9.5 Ib$^{2}$) &   $\blacktriangle$ \\
      W1050 &  16:47:06.77 & -45:49:55.35  &  29.3  &  4.66  &   3.75  &   O9.5 II$^{2}$   &  $\blacktriangle$  \\
      W1051 & 16:47:07.00 & -45:49:40.20 &     30.5 &               5.10 &      3.45 &                    O9 III$^{2}$ &                    \\
      W1052 & 16:47:07.00 & -45:52:56.05 &       27.5 &               4.17 &  4.00 &     B0-0.5 III (O9 III$^{2}$) &                    \\
      W1056 & 16:47:08.69 & -45:51:01.72 &        30.8 &               5.07 &      3.60 &    O9.5 II$^{2}$ &   $\blacktriangle$ \\
      W1057 & 16:47:08.69 & -45:50:47.18 &       30.2 &               5.15 &      3.45 &      O9.5 II-III (O9.5-B0 Iab$^{2}$) &                  \\
      W1058 & 16:47:08.90 & -45:51:24.65 &       30.3 &               4.96 &      3.42 &      O9.5 I-II (O9 III$^{2}$) &                    \\
      W1060 & 16:47:09.20 & -45:50:48.45 &       31.5 &               4.97 &      3.65 &               O9.5 II$^{2}$ &   $\blacktriangle$ \\
      W1061 & 16:47:09.76 & -45:50:40.33 &      29.5 &               4.47 &      3.60 &         B0-0.5 II-III (O9.5 III$^{2}$) &   $\triangle$   \\
      W1063 & 16:47:10.75 & -45:49:47.95 &      30.7 &               4.84 &      3.70 &      O9.5 III$^{2}$ &   $\blacktriangle$ \\
      W1064 & 16:47:11.51 & -45:49:59.94 &       31.5 &               4.92 &      3.65 &    O9.5 I-II (O9.5 Iab$^{2}$) &                    \\
      W1065 & 16:47:11.60 & -45:49:22.54 &       26.5 &               5.46 &      3.45 &    B0 Ib$^{2}$ &   $\blacktriangle$ \\
      W1067 & 16:47:13.38 & -45:49:10.69 &       23.5 &               5.31 &      3.00 &         B0 Iab$^{2}$ &   $\blacktriangle$ \\
      5940105487399473664$^{3}$ & 16:47:00.37 & -45:53:49.41 &               26.5 &               3.96 &      3.75 &                   B0-0.5 V &                    \\
      5940106488128732416$^{3}$ & 16:46:47.01 & -45:51:47.42 &               28.2 &               4.24 &      3.75 &                     B0-0.5 V &         \checkmark \\
 G739$^{1}$ & 16:47:03.51 & -45:51:59.58 &         29.0 &               4.51 &      3.60 &     B0-0.5 II-III &                \checkmark \\
 G787$^{1}$ & 16:47:03.04 & -45:51:55.51 &        28.0 &               4.24 &      4.15 &      O9-O9.5 V  &     \checkmark   \\
 G905$^{1}$ & 16:47:08.76 & -45:51:42.69 &        27.0 &               4.16 &      4.15 &      O9-O9.5 V  &     \checkmark    \\
 G916$^{1}$ & 16:46:57.38 & -45:51:40.42 &        26.0 &               3.94 &      4.15 &      B0-0.5 V  &                    \\
G1002$^{1}$ &   16:47:03.88 & -45:51:31.25   &   31.5  &   5.00    &   3.65   &        O9-9.5 II-III  &                    \\
G1073$^{1}$ & 16:47:00.76 & -45:51:25.07 &       30.2 &               4.62 &      3.60 &      O9-9.5 II-III  &                    \\
G1115$^{1}$ &  16:47:07.22 & -45:51:20.55   &   31.8   &   5.02    &   3.80   &        O9-9.5 II-III  &                    \\
G1352$^{1}$ & 16:47:10.73 & -45:50:58.78 &       29.0 &               4.30 &      3.90 &   O9-O9.5 V  &         \checkmark \\
G1379$^{1}$ & 16:47:13.74 & -45:50:52.54 &       28.4 &               4.18 &      4.15 &     B0-0.5 II-III  &          \\
G1461$^{1}$ & 16:46:58.25 & -45:50:45.87 &       29.8 &               4.33 &      3.85 &   O9-O9.5 V  &        \checkmark  \\
G1686$^{1}$ & 16:47:01.95 & -45:50:21.82 &      30.2 &               4.68 &      3.60 &      O9-9.5 II-III  &         \checkmark \\
G1730$^{1}$ &  16:47:04.26 & -45:50:18.33   & 30.7  &  4.71   &   3.70    &   O9-9.5 II-III  &          \\
G1846$^{1}$ &  16:47:10.47 & -45:50:06.62   &   29.3   &   5.06   &  3.75  &      O9-9.5 II-III  &        \\
G1858$^{1}$ & 16:47:08.47 & -45:50:05.39 &        28.5 &               4.25 &      4.15 &     B0-0.5 V  &         \checkmark \\

G1916$^{1}$ & 16:47:04.69 & -45:49:59.86 &       29.0 &               4.32 &      3.60 &    B0-0.5 II-III  &       \\
G1967$^{1}$ & 16:47:05.67 & -45:49:55.10 &       28.3 &               4.22 &      4.00 &   O9-9.5 V  &         \checkmark \\
G2007$^{1}$ & 16:47:03.94 & -45:49:50.70 &       30.2 &               4.79 &      3.45 &      O9-9.5 II-III  &                    \\
G3022$^{1}$ & 16:47:01.92 & -45:51:40.20 &      28.0 &               4.29 &      3.75 &      B0-0.5 II-III  &                    \\
G3103$^{1}$ & 16:47:01.48 & -45:51:32.17 &      24.0 &               3.87 &      3.75 &        B1 V  &      \checkmark    \\
G3151$^{1}$ & 16:47:03.32 & -45:51:25.58 &      29.0 &               4.34 &      3.75 &     B0-0.5 II-III  &       \\
G3225$^{1}$ & 16:47:11.79 & -45:51:14.25 &      25.0 &               3.96 &      4.20 &       B1 V  &      \checkmark  \\
G3277$^{1}$ &   16:47:03.62 & -45:51:07.17   &   31.5   &  5.02   &  3.65  &       O9-9.5 II-III  &                    \\
G3330$^{1}$ & 16:47:12.26 & -45:50:59.29 &      28.4 &               4.23 &      4.15 &   O9-9.5 V &         \\
G4020$^{1}$ & 16:47:10.02 & -45:49:17.83 &     29.8 &               4.47 &      4.00 &     B0-0.5 III-V  &         \checkmark \\
G4090$^{1}$ & 16:46:59.60 & -45:49:07.43 &     27.0 &               4.01 &      4.00 &     B0-0.5 V  &                    \\
G4232$^{1}$ &   16:47:06.74 & -45:50:57.79   &   29.5   &   5.22       &  3.45  &        O9-9.5 I-II  &                    \\
G4236$^{1}$ & 16:47:03.83 & -45:50:54.01 &       29.0 &               4.25 &      3.45 &        B0-0.5 II-III  &                    \\
G4267$^{1}$ & 16:47:09.22 & -45:50:18.13 &       29.2 &               4.30 &      3.85 &     O9-9.5 V  &         \checkmark \\
G4567$^{1}$ & 16:47:03.21 & -45:52:02.73 &      28.4 &               4.27 &      4.15 &      B0-0.5 II-III &         \checkmark \\
G4657$^{1}$ & 16:47:04.12 & -45:51:43.38 &      30.0 &               4.24 &      4.15 &         B0-0.5 V &    \checkmark    \\
G4667$^{1}$ & 16:47:00.03 & -45:51:43.06 &        26.4 &               4.13 &      4.00 &      B0-0.5 V &   \checkmark     \\
G4675$^{1}$ & 16:47:00.94 & -45:51:41.33 &        24.0 &               3.86 &      3.75 &         B1 V  &    \checkmark     \\
G4679$^{1}$ & 16:47:05.44 & -45:51:40.42 &       30.1 &               4.37 &      3.85 &       O9-9.5 II-III  &         \checkmark \\
G4738$^{1}$ & 16:47:03.57 & -45:51:31.72 &        28.8 &               4.39 &      4.00 &      B0-0.5 V  &         \checkmark \\
G4766$^{1}$ & 16:47:09.74 & -45:51:24.66 &     28.5 &               4.22 &      3.60 &     B0-0.5 V  &     \checkmark   \\
G4862$^{1}$ & 16:47:04.70 & -45:51:09.37 &        30.2 &               4.47 &      3.75 &    B0-0.5 II-III  &         \checkmark \\
G4882$^{1}$ & 16:47:04.77 & -45:51:06.26 &           30.2 &               4.86 &      3.75 &    O9-9.5 II-III  &          \\
G5049$^{1}$ &  16:47:08.95 & -45:50:29.57   &   29.0   &  4.37           &   3.75    &    O9-9.5 V  &        \\
G5055$^{1}$ & 16:47:10.65 & -45:50:27.38 &    29.0 &               4.28 &      3.75 &    B0-0.5 V  &         \checkmark \\
G5331$^{1}$ & 16:47:09.08 & -45:49:06.16 &       29.5 &               4.33 &      3.75 &       O9-9.5 V  &         \checkmark \\
G5453$^{1}$ & 16:47:01.34 & -45:51:10.85 &        25.0 &               3.86 &      4.20 &           B0-B0.5 V &         \checkmark \\
G5469$^{1}$ &   16:47:03.37 & -45:51:03.42   &    30.2  &  5.29  &  3.60   &      B0 I-II &          \\
G5748$^{1}$ & 16:47:04.51 & -45:50:55.36 &        28.5 &               4.25 &      3.75 &            B0-0.5 V  &         \checkmark \\
G5777$^{1}$ & 16:47:02.69 & -45:50:25.45 &        28.2 &               4.25 &      3.85 &     B0-0.5 V &         \checkmark \\
G5834$^{1}$ &  16:47:03.01 & -45:50:51.15   &  30.3  &  5.07  & 3.42 &    O9-9.5 II-III &      \\
\end{longtable}%

\refstepcounter{section}
\section*{Appendix~\thesection: Example of CMFGEN spectral fit}
\label{sec:Appendix_modelfit}

\begin{figure}[H]
  \centering
  \begin{subfigure}[b]{\columnwidth}
    \centering
    \includegraphics[width=0.92\textwidth]{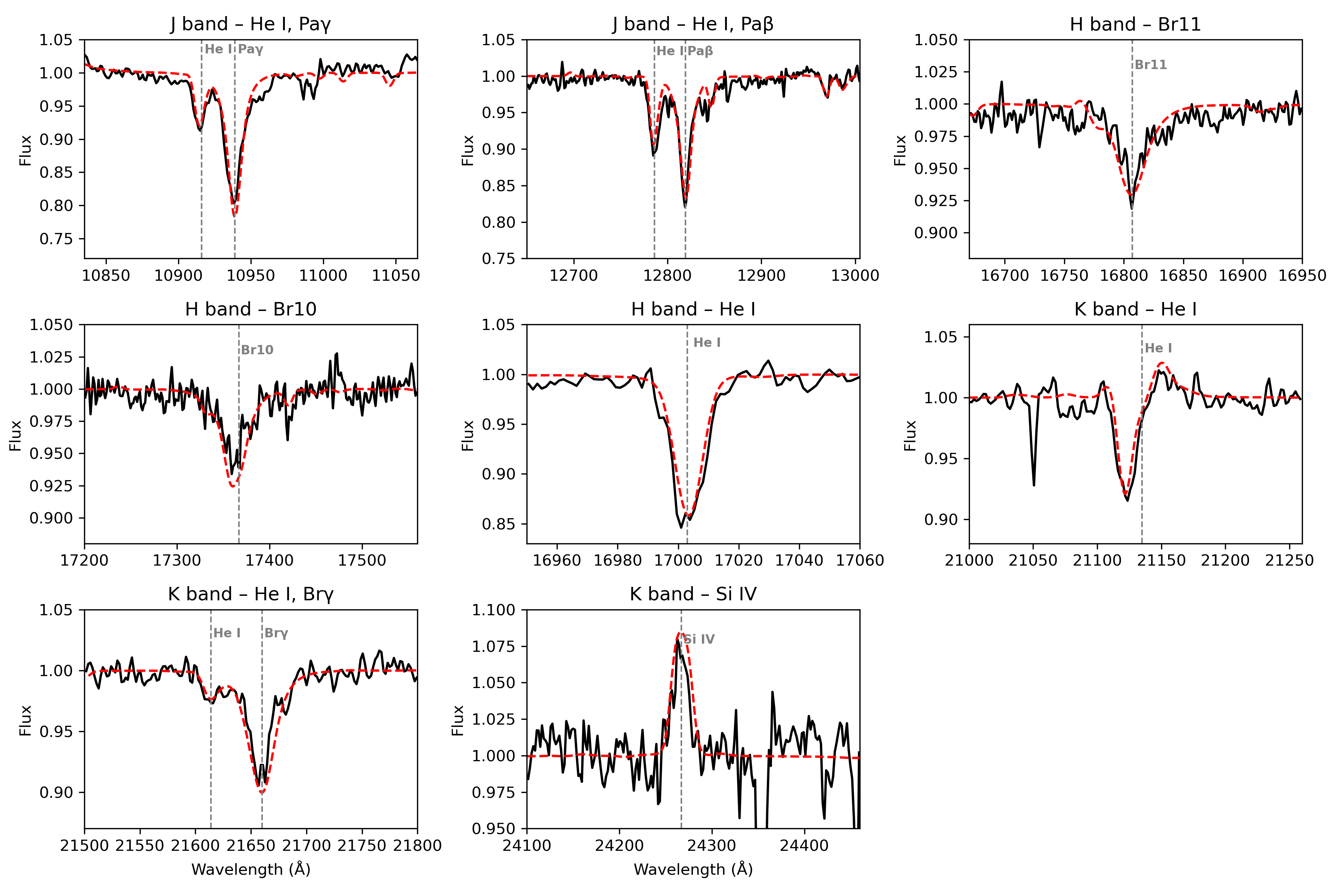}
  \end{subfigure}

  \vspace{2mm}  

  \begin{subfigure}[b]{\columnwidth}
    \centering
    \includegraphics[width=0.92\textwidth]{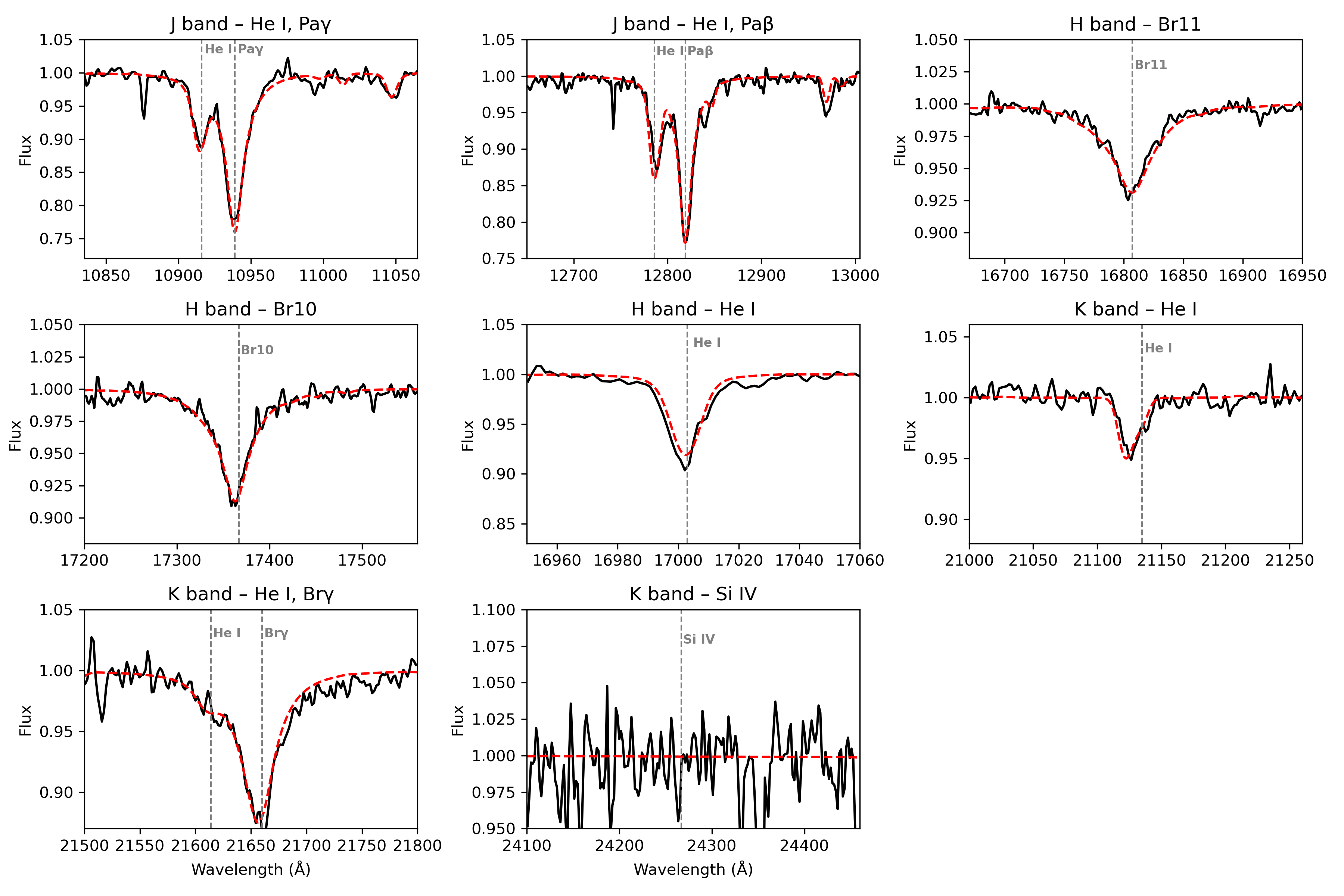}
  \end{subfigure}

  \caption{CMFGEN model fits (red) to KMOS spectra (black) in the $J$, $H$, and $K$ bands.
           \textbf{Top}: star W1051 (O9 III) — $T_\mathrm{eff}=30.5$ kK, $\log g = 3.45$.
           \textbf{Bottom}: star G787 (O9-O9.5 V) — $T_\mathrm{eff}=28$ kK, $\log g = 4.15$.
           Key hydrogen and helium diagnostic lines are indicated in each panel.}
  \label{fig:fit_examples}
\end{figure}

\twocolumn

\section{Impact of adopted parameters on the HRD}
\label{sec:Appendix_parameters}

In this section we examine how sensitive our results are to the adopted distance, extinction law, and stellar rotation and metallicity of the evolutionary models.
In each case, we recomputed the stellar luminosity or re-evaluated the HRD under alternative assumptions in order to ascertain the potential change in inferred age. 

\subsection{Effect of adopted distance}

\begin{figure}[H]
  \centering
  \includegraphics[width=\columnwidth]{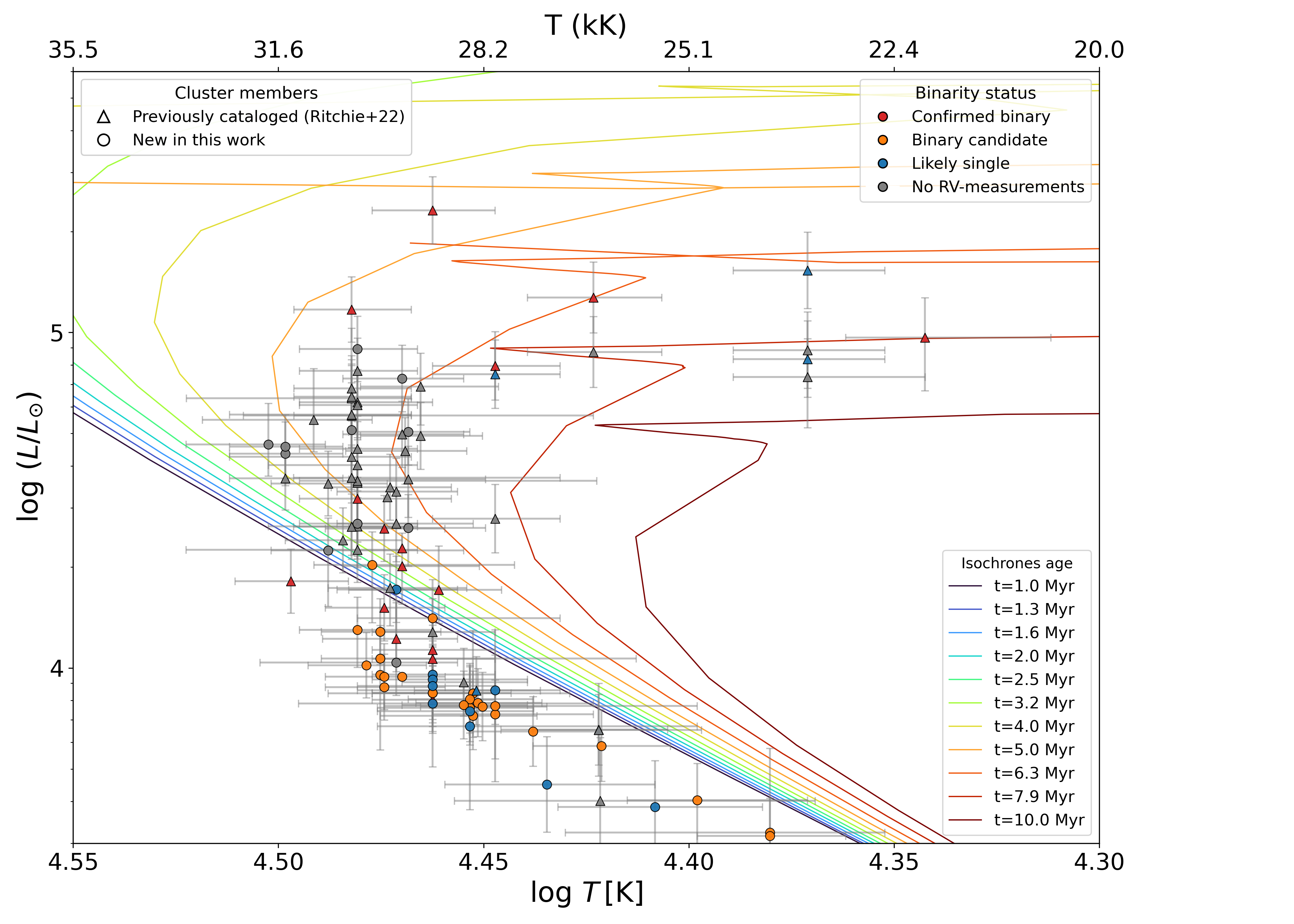}
  \caption{Westerlund~1's HRD, using a hypothetical distance of 2.8 kpc. Lowering the distance to 2.8~kpc decreases stellar luminosities by $\sim$0.3 dex, placing many of the main sequence stars below the youngest available isochrones \citep{Yusof2022}.}
  \label{fig:HRD_distance_appendix}
\end{figure}

The adopted distance affects the luminosities of all stars, as illustrated in Fig.~\ref{fig:HRD_distance_appendix}. 
A significantly shorter distance of $\sim$2.8 kpc, as proposed by \citet{Aghakhanloo2021}, shifts all stars' luminosity downward by approximately 0.3 dex. 
In this scenario, most main-sequence stars would have lower luminosity and effective temperature than predicted even by the youngest available isochrones, making it impossible to reconcile their positions with any standard evolutionary track. 
Similarly, the main sequence region would lie in a regime inconsistent with single-star models, and the evolved supergiants would appear anomalously faint. 
Such a short distance would also imply systematically lower stellar mass-loss rates inferred from radio and sub-mm diagnostics (by a factor of $\sim$0.4), as shown by the ALMA study of Wd1 supergiants by \citet{Fenech2018}.
We therefore conclude that a distance as low as $\sim$3 kpc cannot accommodate either the observed HRD or the derived wind properties of Wd1 members.

\subsection{Analysis with evolutionary isochrones with different metallicity}

\begin{figure}[H]
  \centering
    \includegraphics[width=\columnwidth]{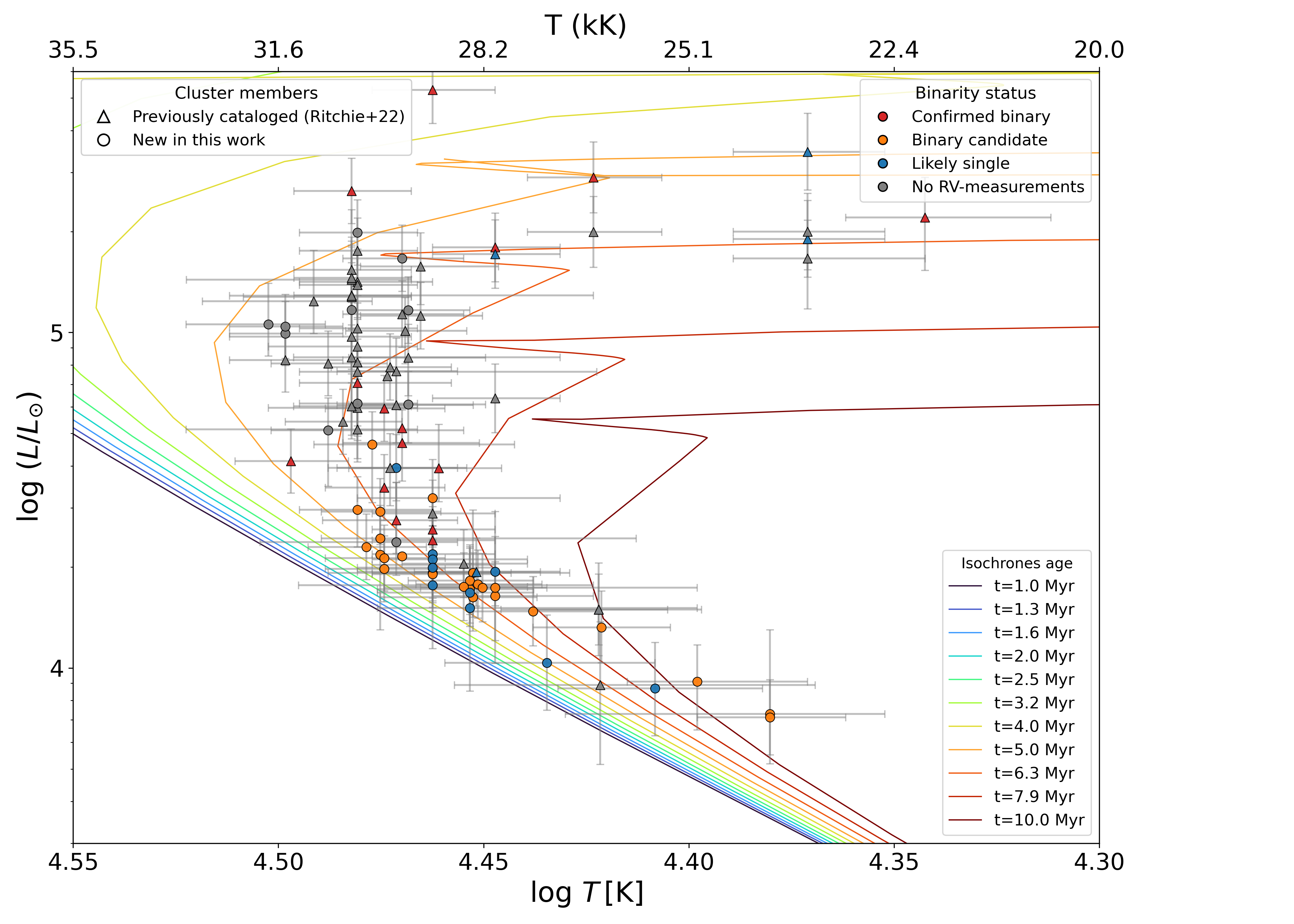}
  \caption{HR diagram comparing the observed stars with Geneva evolutionary tracks at solar metallicity ($Z=0.014$; $v_{\rm ini}=0$, \citealt{Ekstrom2012}).}
  \label{fig:HRD_metallicity_appendix}
\end{figure}

We tested Geneva stellar evolution models at solar metallicity ($Z=0.014$,  \citet{Ekstrom2012}), as shown in Fig.~\ref{fig:HRD_metallicity_appendix}. 
At a given effective temperature the solar-metallicity isochrones are observed to be 0.1–0.12 dex brighter in $\log L_\ast$ than the $Z=0.020$ tracks.
In terms of cluster age the impact is minor: switching from $Z=0.020$ to $Z=0.014$ alters the best-fit turn-off age by less than 0.5 Myr, although we note that in this scenario the main sequence stars would yield discrepant higher ages. 
We conclude that reasonable uncertainties in metallicity do not significantly affect the derived age. 

\subsection{Analysis with evolutionary isochrones with different initial rotational velocity}

\begin{figure}[H]
  \centering
    \includegraphics[width=\columnwidth]{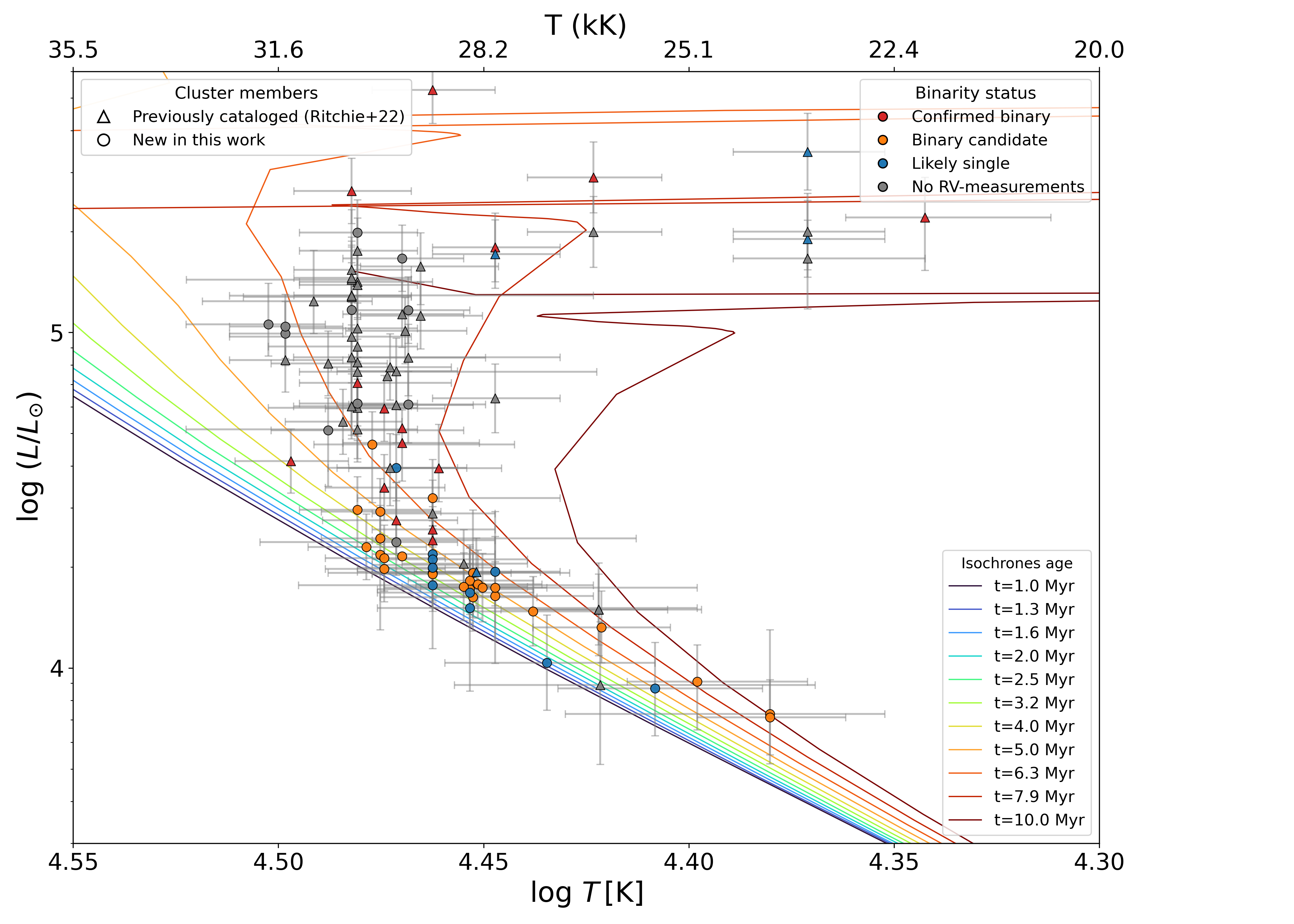}
  \caption{HR diagram showing Geneva tracks ($Z=0.020$, \citep{Yusof2022}) for $v_{\rm ini} = 0.4\,v_{\rm crit}$ rotating models.}
  \label{fig:HRD_rotation_appendix}
\end{figure}

Figure~\ref{fig:HRD_rotation_appendix} compares the location of the sample stars in the HRD against $Z=0.020$ \citep{Yusof2022} isochrones computed with initial angular velocity of \(v_{\rm rot}=0.4\,v_{\rm crit}\).
Rotation increases the main-sequence lifetime through enhanced internal mixing, so the rotating isochrone reaches the observed turn-off temperature after additional \(\simeq0.8\) Myr.  
The turn off would be consistent with a slightly older 6.3-7.9 age, although in this case the main sequence members would point to a younger value.
Moreover, we find no evidence for stars rotating near the critical velocity. 
Consequently, the larger ages implied by near-critical rotational tracks are disfavoured.

\subsection{Effect of adopting a different foreground extinction law}

\begin{figure}[H]
  \centering
    \includegraphics[width=\columnwidth]{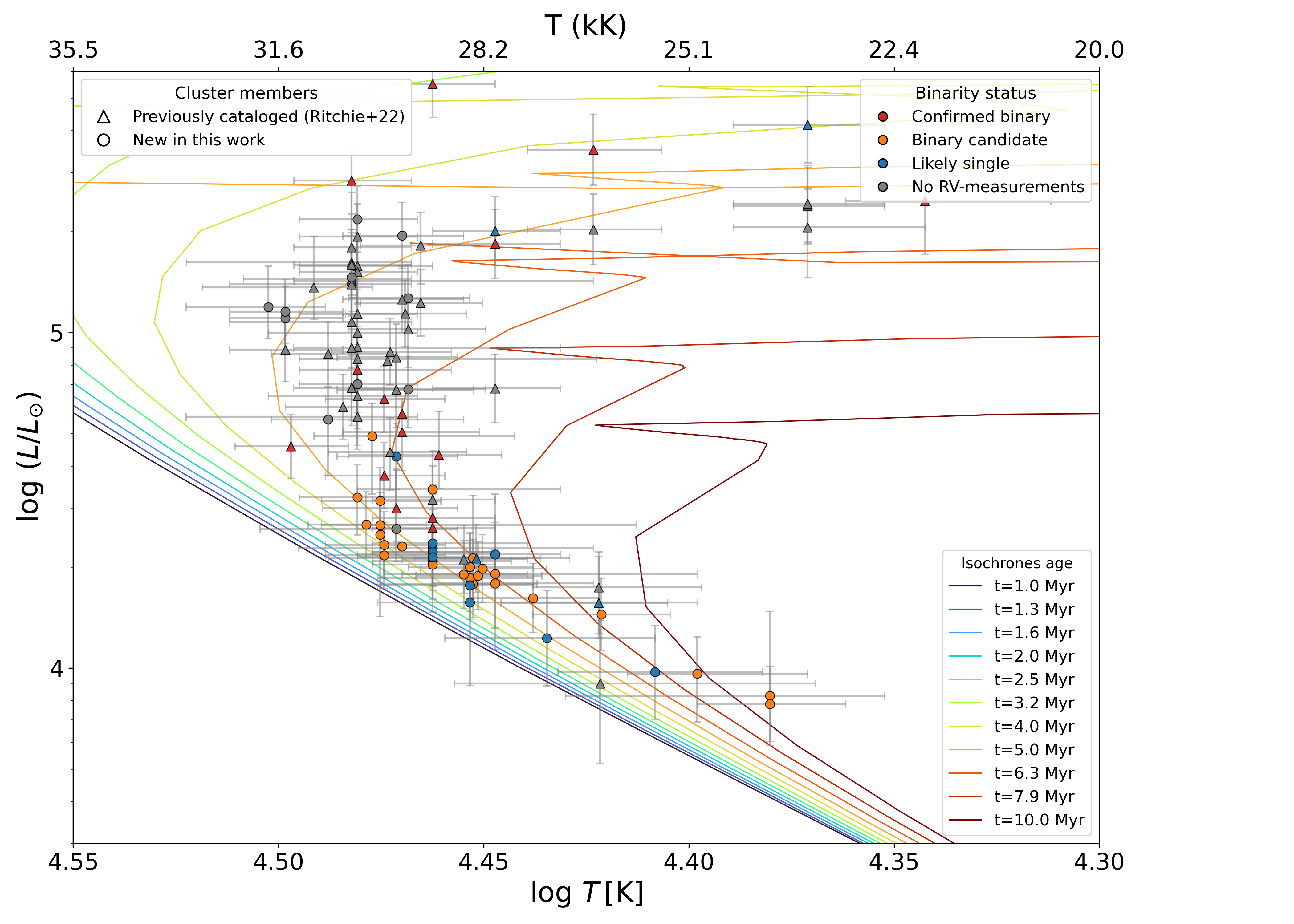}
  \caption{HR diagram with stellar luminosities computed using the extinction law of \citet{Damineli2016}. 
  The non-rotating Geneva isochrones ($Z=0.020$; \citealt{Yusof2022}) are shown for comparison.}
  \label{fig:HRD_extinction_appendix}
\end{figure}

We adopted the \citet{Hosek2018} extinction curve to calculate stellar luminosities, which was derived for environments like Wd1 and the Galactic Centre.
An alternative approach is to consider the steeper optical–IR law from \citet{Damineli2016}. 
Using the Damineli law would make our stars slightly more luminous by up to $\Delta \log L \sim0.05$ dex on average (Figure~\ref{fig:HRD_extinction_appendix}). 
We found that this negligibly impacts the derived cluster age.
A uniform increment of 0.05 dex in $L_{\ast}$ would translate into a slightly older age for main sequence stars, while the turn off and evolved stars would point to younger ages.
We conclude that using a different extinction law does not alter our results significantly. 

\end{appendix}

\end{document}